\documentclass[a4paper,11pt]{article}
\pdfoutput=1 % if your are submitting a pdflatex (i.e. if you have
\usepackage{jcappub} % for details on the use of the package, please
\usepackage[T1]{fontenc} % if needed
\usepackage{ulem}

\usepackage[outline]{contour}
\usepackage{xcolor,graphicx}

\renewcommand{\a}{\alpha}
\renewcommand{\b}{\beta}

\renewcommand{\ss}{\sigma}

\newcommand{\beq}{\begin{equation}}
\newcommand{\eeq}{\end{equation}} %\indent}
\newcommand{\eei}{\end{equation}\indent\indent}
\newcommand{\bc}{\begin{center}}
\newcommand{\ec}{\end{center}}
\newcommand{\bea}{\begin{eqnarray}}
\newcommand{\eea}{\end{eqnarray}}
\newcommand{\ba}{\begin{array}}
\newcommand{\ea}{\end{array}}

\def\case#1/#2{\textstyle\frac{#1}{#2} }

\newcommand{\fett}[1]{\boldsymbol{#1}}
\newcommand{\be}{\begin{equation}}
\newcommand{\ee}{\end{equation}}

\newcommand{\nab}{\fett{\nabla}}
\newcommand{\dd}{{\rm d}}
\newcommand{\stuff}{\mathcal{H}_0^2\Omega_{\rm m0}}

\newcommand{\PS}{{\rm PS}}
\newcommand{\ST}{{\rm ST}}

\newcommand{\myArxiv}[1]{{\tt \href{http://arxiv.org/abs/#1}{{\color{black}[{\color{blue}#1}]}}} \href{http://inspirehep.net/search?p=find+EPRINT+#1}
 {{\color{black}[{\color{blue} {\small in}SPIRE}]}}}
\newcommand{\inspire}[1]{\href{http://inspirehep.net/search?p=find+J+#1}
 {{\color{black}[{\color{blue} {\small in}SPIRE}]}}}
\newcommand{\book}[1]{\href{http://inspirehep.net/search?p=#1}
 {{\color{black}[{\color{blue} {\small in}SPIRE}]}}}
\newcommand{\DOI}[2]{ 
 \href{http://dx.doi.org/#1}{\color{blue} #2}
}
\newcommand{\inspired}[1]{\href{http://inspirehep.net/search?p=#1}
 {{\color{black}[{\color{blue} {\small in}SPIRE}]}}}

\definecolor{darkgreen}{rgb}{0,0.5,0}

\contourlength{0.4pt} % adjusts boldness by using \contour{black}{ text }
\contournumber{10} % ... by reprinting it with n-times with an offset

\title{
Relativistic perturbations in 
 \contour{black}{$\Lambda$}CDM: 
 Eulerian \& Lagrangian approaches
}

\author{Eleonora Villa}
\author{and Cornelius Rampf}

\affiliation{Institute of Cosmology and Gravitation, University of Portsmouth,\\
Dennis Sciama Building, Burnaby Road, Portsmouth PO1 3FX, United Kingdom}

\emailAdd{eleonora.villa@port.ac.uk}
\emailAdd{cornelius.rampf@port.ac.uk}

\abstract{
We study the relativistic dynamics of a pressure-less and irrotational 
fluid of dark matter (CDM) with a cosmological constant ($\Lambda$), up to 
second order in cosmological perturbation theory. In our analysis we also
account for vector and tensor perturbations and include primordial non-Gaussianity.
We consider three gauges: the synchronous-comoving gauge, 
the Poisson gauge and the total matter gauge, where the first is the unique relativistic Lagrangian frame of reference, and the latters are convenient gauge choices for
Eulerian frames. 
Our starting point is the metric and fluid variables in the Poisson gauge up to 
second order. We then perform the gauge transformations to the synchronous-comoving gauge and subsequently to
the total matter gauge. 
Our expressions for the metrics, densities, velocities, and the gauge generators 
are novel and coincide with known results in the limit of a vanishing cosmological constant.
}

\begin{document}
\maketitle
\flushbottom

%%%%%%%%%%%%%%%%%%%%%%%%%%%%%%%%%%%%%%%%%%%%%%%%%%%%%%%%%%%%%%%%%%%%%%%%%%%%%%%%%%%%%%%%%%%%%%%%%%%

\section{Introduction}
%%%%%%%%%%%%%%%%%%%%%%%%%%%%%%%%%%%%%%%%%%%%%%%%%%%%%%%%%%%%%%%%%%%%%%%%%%%%%%%%%%%%%

To put it in simple words, cosmological structure formation is nothing but fluid \mbox{dynamics}. 
There are two frameworks to study fluid dynamics, namely the Lagrangian and the Eulerian frame. 
The Eulerian frame of reference is associated with an observer's coordinate system, where the observer is studying how the distant streams of matter
are clustering. The Lagrangian frame of reference, by contrast, is the physical frame 
attached to the clustering matter elements, i.e., each observer is literally following an individual matter element. 
In the simple Newtonian world, there is only one Eulerian frame, because the infinite Eulerian coordinate systems are physically equivalent, thanks
to the Galilean invariance of the Eulerian fluid equations. In this Newtonian set up, the Eulerian and Lagrangian frame are uniquely defined by the
spatial coordinate transformation embedded in an Euclidean geometry
\be \label{NLPT}
  \bar x^i  = \bar q^i + \bar\Psi^i \,,
\ee
where the bar denotes Euclidean coordinates, $\bar x^i(\bar \tau,\bar{\fett{q}}) $ is the Eulerian coordinate, $\bar{q}^i$ denotes the Lagrangian coordinate 
(i.e., initial position) of a given fluid element, and $\bar\Psi^i(\bar\tau,\bar{\fett{q}}) $ is the Lagrangian displacement field.

In General Relativity (GR), however, this situation is more involved, mainly because of the covariant formulation of 
the relativistic equations of motion. To be more specific,
the \textit{fully relativistic} Lagrangian frame can be uniquely identified with the synchronous-comoving gauge, because the resulting space-time coordinates %, $q^\mu=(\tau,q^i)$, 
are indeed the ones \mbox{attached} to the matter \cite{Ehlers:1993gf}.\footnote{
The locally inertial (or Minkowskian) Lagrangian frame, 
by contrast, is represented by 
the locally free-falling coordinate system along the worldline of the observer 
comoving with the particle, i.e., the Fermi frame, see e.g.~\cite{Manasse:1963zz,Pajer:2013ana}.
}
But there are infinite possibilities for selecting an Eulerian frame in GR. Even defining an Eulerian frame in GR is not so straightforward.
Naturally, perhaps because our
thinking is Newtonian, one could define a generic Eulerian frame in GR with the following correspondence to the Newtonian theory:
\begin{quote}
 \textit{In General Relativity, an Eulerian frame can be identified with a certain gauge if the resulting equations of motion take the classical Newtonian form.}
\end{quote}
As a direct consequence of this definition, it follows that the gauge generator $\xi^\mu$ of the gauge transformation
\be
   x^\mu = q^\mu + \xi^\mu
\ee
is the 4D relativistic counterpart of the 3D Newtonian displacement field (cf.~with eq.\,(\ref{NLPT})). Here, $x^\mu(\eta,x^i)$ are the Eulerian coordinates of a given Eulerian gauge, and $q^\mu= (\tau,q^i)$ are the coordinates of the synchronous-comoving gauge. 
With the above quote, we fix the spatial coordinates of a generic Eulerian frame.
To get a specific Eulerian frame in GR, one then needs to fix the time coordinate, which is in principle arbitrary. One particular convenient choice of the time coordinate 
is certainly to use the proper time of the particles (which is precisely the case in the total matter gauge, see section~\ref{sec:lageul}).

All gauges in GR are physically equivalent, 
but the physical measurement of the density and the velocity of the matter streams depends on the particular Eulerian coordinate system chosen by a given observer.  
For example, convenient Eulerian gauges are the Poisson gauge 
and the total matter gauge, the very
Eulerian gauges we shall investigate. The Poisson gauge is particularly 
 common for studying the gravitational lensing \cite{Seljak:1996is}, whereas the total matter gauge is important for the Eulerian biasing, and when relating GR calculations to investigations in a flat space-time, such as Newtonian $N$-body simulations \cite{Hidalgo:2013mba,Rampf:2014mga,Bertacca:2015mca}.
The synchronous-comoving gauge, apart from being the unique Lagrangian frame of relativistic fluid flow, is the gauge for studying the Lagrangian biasing,
especially when considering the effect of the primordial non-Gaussianity \cite{Villa:2014foa,Matarrese:2008nc,Wands:2009ex,BMPR2010,Bruni:2011ta,Bruni:2014xma}.

In this paper we consider the cosmological fluid in the above mentioned gauges in the framework of cosmological perturbation theory. 
(For different solution techniques, see 
e.g., refs.\,\cite{Rampf:2012pu,Rampf:2013dxa,Rampf:2013ewa,Villa:2014aja,Buchert:2012mb,Buchert:2013qma,Milillo:2015cva,Alles:2015vua}.)
In detail, we study the relativistic dynamics
of a pressure-less and irrotational 
fluid of dark matter (CDM) with a cosmological constant ($\Lambda$), up to 
second order in cosmological perturbation theory.
We begin with reporting the results of refs.~\cite{BMR2006,BMPR2010} in the Poisson gauge, simplify their expressions and provide explicit solutions for all time coefficients, and then we obtain the results in the synchronous-comoving gauge and in the total matter gauge by means of gauge transformations.
 We generalise to a $\Lambda$CDM Universe with primordial non-Gaussianity the results for the synchronous-comoving and 
the Poisson gauge of ref.~\cite{MMB}, which were
restricted to an Einstein--de Sitter (EdS) model, and extend the analysis to
the total matter gauge. In this paper, almost all of the
results are new, only few results were obtained before in the literature.
For example, the matter density in the synchronous-comoving gauge was obtained in
refs.~\cite{BMPR2010,Uggla:2013kya,Uggla:2014hva,BHMW14,Rampf:2014mga}, whereas the 
matter density in the total matter gauge was obtained in refs.~\cite{Rampf:2014mga,BHMW14,Uggla:2013kya,Uggla:2014hva}. In the literature, there are no known $\Lambda$CDM results for the second-order metric in the synchronous-comoving and for the total matter gauge, except for ref.~\cite{Rampf:2014mga} for scalar and vector modes, however obtained with the gradient expansion technique. Here we find the full metric results, the densities and the velocities for these three gauges.
In our analysis, we pay particularly attention to distinguish between
Newtonian-like and GR-like contributions.
Note that the fluid approximation and thus our approach is only valid on sufficient large scales, i.e.,  before matter streams start to intersect (``shell-crossing'').
Furthermore, our analysis is restricted to the fastest growing mode solutions, i.e., we discard decaying modes.

This paper is organised as follows. In the following section we introduce our 
conventions of the metric, density and velocity perturbations. 
In section~\ref{gaugetformulas}
we give the general formulas for gauge transformations up to second order.
Then in~\ref{sec:lageul} we give the definitions of the used gauges.
The derivation of our results in the Poisson gauge are given in section~\ref{Poisson}. 
This analysis serves as our basis to perform the gauge transformation to 
the synchronous-comoving
gauge (section~\ref{sec:PoissonToSynch}). In section~\ref{synchro2x} we then transform
the Lagrangian results to the total matter gauge.
The former two sections are intentionally kept quite pedagogical. For the readers interested only in the results we refer to the compact section~\ref{summary}. 
Finally, we conclude in~\ref{conclude}.

Notation: we use $\mu, \nu ...$ for space-time indices and $i, j, ...$ for spatial indices. Summation over repeated indices is assumed. If not otherwise stated, spatial indices are raised and lowered with the Kronecker delta. A comma denotes spatial partial derivative and a dot denotes partial derivative with respect to the conformal time.
A bar denotes from here on FLRW background quantities. We set $c=1$.
Quantities in the synchronous-comoving gauge, in the Poisson gauge, and in the total matter gauge are respectively denoted with the subscripts \mbox{S, P, and T.} Consistently, quantities which arise because of a gauge transformation are abbreviated with the respective double subscripts, e.g., the gauge generator for the transformation synchronous-comoving to the total matter gauge is denoted with $\xi^\mu_{\ST}$.
For an overview of our used notation see tab.~\ref{tab:notation} on page~\pageref{tab:notation}.

%%%%%%%%%%%%%%%%%%%
\section{The FLRW metric perturbed up to second order}\label{sec:FLRW}

The components of a spatially flat FLRW metric perturbed up to second order are written as
\begin{subequations}
   \label{metricIIany}
\begin{align} 
    g_{00}&=-a^2\left( 1+2 \psi_1+\psi_2 \right) \\
    g_{0i}&=a^2\left( B_{1\,i}+\frac{1}{2} B_{2\,i}\right) \\
    g_{ij}&=a^2\left[ \delta_{ij}+2C_{1\,ij}+C_{2\,ij}\right] \,,
\end{align}
\end{subequations}
where the background part (which is by definition only time-dependent) is given by
\begin{equation} \label{scalef}
\overline{g}_{00}=-a^2(\eta) \,, \qquad\qquad \overline{g}_{ij}=a^2(\eta)\,\delta_{ij} \,.
\end{equation}
Here, $\eta$ is the conformal time ($a\, \dd \eta = \dd t$, where $t$ is the cosmic time), and $a(\eta)$ is the FLRW scale factor, which obeys the Friedmann equation (see section~\ref{sec:0}).
It is convenient to split the perturbations into the so-called scalar, 
vector and tensor parts,
\begin{eqnarray}
B_{(r)\,i}&=&\partial_i B_{(r)}+\omega_{(r)\,i} \label{vec}\\
C_{(r)\,ij}&=&-\phi_{(r)}\delta_{ij}+D_{ij}E_{(r)}+\frac{1}{2}\partial_i F_{(r)\,j}
+\frac{1}{2}\partial_j F_{(r)\,i}
+\frac{1}{2}\chi_{(r)\,ij}\,, \label{tens}
\end{eqnarray}
where $(r)= 1, 2$, stand for the $r$th order of the perturbation, we make use of the operator $D_{ij}\equiv\partial_i\partial_j-(1/3)\nab^2\delta_{ij}$, and $\omega_{(r)\, i}$ and $F_{(r)\,i}$ are transverse vectors, i.e. 
$\partial^i\omega_{(r)\,i}=\partial^i F_{(r)\,i}=0$ and $\chi_{(r)\,ij}$ 
is a transverse and trace-free tensor, i.e.,
$\partial^i\chi_{(r)\,ij}=0$, $\chi^{i}_{(r)\,i}=0$. 
Historically, the reason why such a splitting was introduced \cite{Bardeen:1980kt,kodama} is that at first order these different perturbation modes are decoupled
from each other in the perturbed evolution equations, so that they can be studied separately.
This property does not hold anymore beyond the linear regime, where higher-order perturbations are sourced by products of lower-order perturbations.  Of course, despite the fact that different perturbation modes generally 
do not decouple at higher orders, the above splitting is still physically meaningful.

So far our considerations are fairly general. In the case of the $\Lambda$CDM Universe,
the metric in \eqref{metricIIany} can be simplified, as we can consistently neglect first-order vector and tensor perturbations: linear vector perturbations are not produced by standard inflationary scenarios and in any case decay with the expansion of the Universe \cite{MMB}, and tensor perturbations are believed to give a negligible contribution to the matter dynamics.
However, we should keep in mind that the same reasoning does not apply to second-order 
perturbations: in the non-linear case, scalar, vector and tensor modes are coupled and the second-order vector and tensor 
contributions are generated by first-order scalar perturbations 
even in the absence of linear vector and tensor perturbations, \cite{MMB}.
With these simplifications, 
the spatially flat FLRW metric perturbed up to second order is
\begin{subequations}
 \label{metricICany}
\begin{align}
g_{00} &= -a^2\left( 1+2 \psi_1+\psi_2 \right) \\
g_{0i}&= a^2\left( \partial_i B_1+\frac{1}{2}\, 
\partial_i B_2+\frac{1}{2}\, \omega_{2\,i} \right) \\
g_{ij} &= a^2\left[
\left( 1 -2 \phi_1 - \phi_2 \right)\delta_{ij}+
D_{ij}\left( 2 E_1 +E_2 \right)+\frac{1}{2}\partial_i F_{2\,j}+\frac{1}{2}\partial_j F_{2\,i}+\frac{1}{2}\chi_{2\,ij}\right] \,.
\end{align}
\end{subequations} 
The four-velocity of matter is $u^\mu=  {\rm d}x^\mu / {\rm d}\tau$, where $\tau$ is the proper (comoving) time, comoving with the fluid. For latter convenience we set
\begin{equation} \label{umu}
u^\mu=\frac{1}{a}\left(\delta^\mu_0+v^\mu\right),
\end{equation}
where $v^\mu = v_1^\mu + v_2^\mu/2 + \ldots$ is the peculiar velocity (peculiar in the spatial and temporal sense).
From the normalisation condition $u^\mu u^\nu g_{\mu\nu}=-1$ we 
obtain the constraint for the time component of $v^\mu$, which reads up to second order (in any gauge)
\begin{equation} \label{v0any}
v^0 =  -\psi_1-\frac{1}{2}\psi_2+\frac{3}{2}\psi_1^2+\frac{1}{2}v_{1k}v_1^k+v_{1k}B_1^{,k}\,.
\end{equation}
The perturbations of the spatial components $v^i$ split as usual in scalar and vector parts
\begin{equation} \label{velocity}
 v^i _{(r)}= \delta^{ij}v_{(r)\,,j}+w_{(r)}^i\,,
\end{equation}
where $\partial_iw_{(r)}^i=0$.
Finally, the perturbation in the matter density is written as $\rho=\overline{\rho}+\rho_1+\rho_2/2+\ldots$, where the background density $\overline{\rho}$ is time-dependent only. In this paper we work with the density contrast which is defined by
\be \label{def:delta}
  \delta(\eta, \fett{x}) \equiv \frac{\rho(\eta,\fett{x})  -\bar \rho(\eta) }{\bar \rho (\eta)}  \,,
\ee
and is expanded as
\be \label{perturbedDelta}
  \delta = \delta_1 + \frac{\delta_2}{2} 
\ee
up to second order.

%%%%%%%%%%%%%%%%%%%%%%%%%%
\section{Gauge transformations in perturbation theory} \label{gaugetformulas}

The theory of the gauge transformations in any given background space-time and beyond the linear order was applied to cosmology in refs.~\cite{MBgeo,flanagan}, following the approach of refs.~\cite{stewart74,Sonego:1997np}. See \cite{malik&wands} for a recent review. 

We will adopt in the following the so-called passive approach, where a gauge transformation is seen as a coordinate transformation $x^\mu \rightarrow \tilde{x}^\mu$. Up to second order it reads
\begin{align} \label{TdG}
  \tilde x^\mu (x^\a) &= x^\mu - \xi^\mu_{1}(x^\alpha) + \frac{1}{2} \left[ \xi^\mu_{1,\nu}(x^\alpha) \,\xi^\nu_{1}(x^\alpha) - \xi^\mu_{2}(x^\alpha) \right] \,,
\intertext{with its inverse} 
 \label{TdGi}
  x^\a(\tilde x^\mu) &= \tilde x^\a + \xi^\a_{1}(\tilde x^\mu) + \frac{1}{2} \left[ \xi^\a_{1,\b}(\tilde x^\mu)\, \xi^\b_{1}(\tilde x^\mu) + \xi^\a_{2}(\tilde x^\mu) \right]  \,,
\end{align}
where all the quantities are evaluated at the same point on the background space-time where the coordinates $x^\mu$ and $\tilde x^\mu$ coincide. As usual, the four vectors $\xi_{(r)}^\mu$ can be decomposed into scalar and vector parts
\beq \label{xidecomp}
\xi^0_{(r)} = \alpha_{(r)}\, , \qquad \xi^{i}_{(r)} = \partial^i \beta_{(r)} + d^i_{(r)}\, , \qquad  \text{with} \quad \partial_i d^i_{(r)} = 0\,.
\eeq
In order to find the components of a tensor in the new coordinates we simply start from the standard transformation rule for the metric tensor
\begin{equation}
\tilde{g}_{\mu\nu}(\tilde x^\a)=\frac{\partial x^\sigma}{\partial \tilde x^\mu}\frac{\partial x^\lambda}{\partial \tilde x^\nu}g_{\sigma\lambda}(x^\a(\tilde x^\rho))\,,
\end{equation} 
and expand the Jacobian matrix, $\frac{\partial x^\sigma}{\partial \tilde x^\mu}$, and the argument of the metric components on the r.h.s. up to second order. 
The result of this infinitesimal coordinate transformation is naturally expressed by means of the Lie derivative,\footnote{The explicit expressions for the Lie derivative %for vectors, 1-forms and the metric tensor 
can be found in appendix \ref{lieapp}.}  $\mathcal{L}_{\xi}$. 
%as long as the map is invertible. 
We have up to second order
\begin{equation}
\tilde{g}_{\mu\nu}=g_{\mu\nu}+\mathcal{L}_{\xi_1} g_{\mu\nu}+\frac{1}{2}\left(\mathcal{L}^2_{\xi_1}g_{\mu\nu}+\mathcal{L}_{\xi_2}g_{\mu\nu}\right).
\end{equation} 
In general, by expanding the standard transformation rules under the coordinate transformation \eqref{TdG} for any unexpanded quantity $\mathbf{T}$ --- a scalar, a vector, or a tensor ---, we find that its change up to second order is given by
\begin{equation}\label{expt}
\tilde{\mathbf{T}}= \mathbf{T}+ \mathcal{L}_{\xi_1}\mathbf{T}+\frac{1}{2}\left(\mathcal{L}^2_{\xi_1}\mathbf{T}+\mathcal{L}_{\xi_2}\mathbf{T}\right) \,.
\end{equation}
In order to find how the perturbations of a given tensor field transform, we consider now the expansion up to second-order of a generic tensor field $\mathbf{T}$ in two different gauges,  
\begin{equation}\label{expq}
  \mathbf{T}= \overline{\mathbf{T}}+ \mathbf{T}_1+\frac{1}{2}\mathbf{T}_2 \,,   \qquad\text{and} \qquad
 \tilde{\mathbf{T}}= \overline{\mathbf{T}}+ \tilde{\mathbf{T}}_1+\frac{1}{2} \tilde{\mathbf{T}}_2  \,.
\end{equation}
The transformation rule for the perturbations is easily obtained by plugging \eqref{expq} in \eqref{expt} and collecting the terms of the same order. It reads
\beq
 \overline{\mathbf{T}}+\tilde{\mathbf{T}}_1+\frac{1}{2} \tilde{\mathbf{T}}_2  =\overline{\mathbf{T}}+  \mathbf{T}_1+\mathcal{L}_{\xi_1}\overline{\mathbf{T}}+\frac{1}{2}\left( \mathbf{T}_2 +2\mathcal{L}_{\xi_1} \mathbf{T}_1+\mathcal{L}^2_{\xi_1}\overline{\mathbf{T}}+\mathcal{L}_{\xi_2}\overline{\mathbf{T}}\right) \,.
\eeq
This relation tells us how to obtain the perturbations in one gauge from the corresponding perturbations in the other gauge, that is
\begin{align} 
  \label{pert1}
 \tilde{\mathbf{T}}_1 &=\mathbf{T}_1+\mathcal{L}_{\xi_1}\overline{\mathbf{T}} 
\intertext{at first order, and}
 \label{pert2}
 \tilde{\mathbf{T}}_2 &= \mathbf{T}_2 +2\mathcal{L}_{\xi_1} \mathbf{T}_1+\mathcal{L}^2_{\xi_1}\overline{\mathbf{T}}+\mathcal{L}_{\xi_2}\overline{\mathbf{T}}
\end{align}
at second order. The r.h.s's of these equations are automatically written in the new gauge.

In the following two subsections, we give the transformation rules for the quantities we are interested in, namely the metric tensor, the density contrast and the four-velocity. They are straightforwardly obtained from eqs.~\eqref{pert1} and~\eqref{pert2} and by the use of the expressions in appendix~\ref{lieapp}.

\subsection{First-order transformations}\label{generalFO}
\paragraph{Metric tensor}
We find the following first-order transformations for the metric tensor: 
\begin{itemize}
\item scalar perturbations
\bea
\tilde \psi_1 &=& \psi_1 + {\cal H} \alpha_1 + \dot \alpha_1  \label{psi1}\\ 
\tilde \phi_1 &=& \phi_1 - {\cal H} \alpha_1 -\frac{1}{3}\nab^2\beta_1 \label{phi1}\\ 
\tilde B_1 &=& B_1 - \alpha_1 + \dot \beta_1 \label{B1}\\
\tilde E_1 &=& E_1 + \beta_1 \,, \label{E1}
\eea
\item vector perturbations
\bea
\tilde \omega_{1i} &=& \omega_{1i} +\dot d_{1i} \\\label{omega1}
\tilde F_{1i} &=& F_{1i} +  d_{1i} \,, \label{F1} 
\eea
\item tensor perturbations
\bea
\tilde \chi_{1 ij} &=& \chi_{1 ij}\, ,\label{chi1} 
\eea
\end{itemize}
where the dot denotes partial derivative with respect to conformal time and ${\mathcal H}=\dot{a}/a= a H$ is the conformal Hubble parameter. 
Note that, since we neglect first-order vector and tensor perturbations, the first-order vector part of the spatial transformation $d_1^i$ is constant and can be set to zero.

\paragraph{Three-velocity} 
The transformation of the temporal part of the peculiar velocity is obtained from eq.\,\eqref{v0any} and reads
\begin{equation}\label{v01}
\tilde{v}^0_1=v^0_1- {\cal H} \alpha_1 - \dot \alpha_1\,.
\end{equation}
For the scalar and vector part of the spatial peculiar velocity we find
\begin{equation}\label{v1}
\tilde{v}_1=v_1-\dot \beta_1 
\end{equation}
and
\begin{equation}\label{w1}
\tilde{w}^i_1=w^i_1-\dot d^i_1\,. 
\end{equation}

\paragraph{Matter density}
Finally, the perturbation of the density contrast transforms as
\begin{equation}\label{rho1}
\tilde{\delta}_1=\delta_1 - 3 {\cal H}\,\alpha_1\,.
\end{equation}

\subsection{Second-order transformations}\label{secGauge}

In order to write in compact form the second-order gauge transformations we collect the contributions from products of first-order transformations defining $\Pi, \Sigma_{i}, \Omega^{i}$ and $\Upsilon_{ij}$. Their explicit expressions are given in Appendix \ref{app:FOsquaredterms}.

\paragraph{Metric tensor}
The second-order transformations for the metric are:
\begin{itemize}
\item scalar perturbations
\bea
\tilde \psi_2 &=& \psi_2 + {\cal H} \alpha_2 + \dot \alpha_2 + \Pi \label{psi2} \\ 
\tilde \phi_2 &=& \phi_2 - {\cal H} \alpha_2 - \frac{1}{6} \Upsilon^k_{\ k} -\frac{1}{3}\nab^2\beta_2 \label{phi2}\\ 
\tilde B_2 &=& B_2 - \alpha_2 + \dot \beta_2 + \nab^{-2} \Sigma^{\phantom{k},k}_{k} \label{B2} \\
\tilde E_2 &=& E_2 + \beta_2 + \frac{3}{4} \nab^{-2} \nab^{-2} \Upsilon^{\phantom{ij},ij}_{ij} - \frac{1}{4} \nab^{-2} \Upsilon^{k}_{\ k} \,, \label{E2} 
\eea 
\item \vskip-0.1cm vector perturbations
\bea
\tilde \omega_{2i} &=& \omega_{2i} + \dot d_{2i} +\Sigma_i - \nab^{-2} \Sigma^{\phantom{k},k}_{k,i} \label{omega2} \\
\tilde F_{2i} &=& F_{2i} + d_{2i} + \nab^{-2} \Upsilon_{ik}^{\phantom{ik},k} - \nab^{-2} \nab^{-2} \Upsilon^{\phantom{kl},kl}_{kl,i}, \label{F2} 
\eea
\item \vskip-0.1cm tensor perturbations
\bea
\tilde \chi_{2ij} &=& \chi_{2ij} + \Upsilon_{ij} + \frac{1}{2} \left( \nab^{-2} \Upsilon^{\phantom{kl},kl}_{kl} - \Upsilon^k_{\ k}\right) \delta_{ij} + \frac{1}{2} \nab^{-2} \nab^{-2} \Upsilon^{\phantom{kl},kl}_{kl, ij}+ \nonumber \\
&& + \frac{1}{2} \nab^{-2} \Upsilon^k_{\ k,ij} - \nab^{-2} \left( \Upsilon_{ik,j}^{\phantom{ik},k}+ \Upsilon_{jk,i}^{\phantom{jk},k}\right)\, , \label{chi2}
\eea
\end{itemize}
where $\nab^{-2}$ stands for the inverse of the Laplacian operator with Euclidean metric.

\paragraph{Four-velocity}
From the normalisation condition \eqref{v0any} in the new gauge we find the transformation of the temporal part of the peculiar velocity which reads
\begin{eqnarray} \label{v02}
\tilde{v}^0_2&=&v^0_2- {\cal H} \alpha_2 - \dot {\alpha}_2+\alpha_1\left[2\left(\dot{v}^0_1-{\cal H}v^0_1\right)+\alpha_1\left({\cal H}^2-\dot{{\cal H}}\right)+{\cal H}\dot {\alpha}_1-\ddot {\alpha}_1\right]\nonumber \\
&&+\xi^i_1\left(2v^0_{1,i}-{\cal H}\alpha_{1,i}-\dot {\alpha}_{1,i}\right)+\dot {\alpha}_1\left(\dot {\alpha}_1-2v^0_1\right)-2\alpha_{1,i}v^i_1+\alpha_{1,i}\dot{\xi}^i_1\,.
\end{eqnarray}
The scalar and vector part of the spatial peculiar velocity are given by
\begin{equation}\label{v2}
\tilde{v}_2=v_2-\dot \beta_2+\nab^{-2}\Omega^k_{\ ,k} %6.28MW
\end{equation}
and
\begin{equation}\label{w2}
\tilde{w}^i_2=w^i_2-\dot d^i_2+\Omega^i-\nab^{-2}\Omega^{k,i}_{\ ,k}\,.%6.29MW
\end{equation}
\paragraph{Matter density}
Finally, the second-order perturbation of the density contrast transforms as
\begin{equation}\label{rho2}
\tilde{\delta}_2=\delta_2- 3 {\cal H} \alpha_2+\alpha_1\left[ \left( 6 {\cal H}^2 + \frac 9 2 \frac{\mathcal H_0^{2}\Omega_{{\rm m}_0}}{a} \right)\,\alpha_1 - 3 {\cal H}\dot{\alpha}_1+2\dot{\delta}_1 - 6 {\cal H} \delta_1 \right]+\left(2\delta_1 - 3{\cal H} \alpha_1\right)_{,k}\xi_1^k\,.
\end{equation}

%%%%%%%%%%%%%%%%%%%%%%%%
\section{Lagrangian gauge and Eulerian gauges}\label{sec:lageul}

The Lagrangian frame for irrotational and pressure-less matter, in the language of gauges, can be associated with the synchronous and comoving gauge (see, e.g., \cite{Ehlers:1993gf,Rampf:2014mga,Bertacca:2015mca}). 
The synchronous gauge is defined by \cite{Landau:1987gn}
\begin{align} 
   B_{\rm S}= \omega_{\rm S}^i &= 0   \,, \qquad (\text{spatial~gauge~condition})  \label{synchS} \\
   \psi_{\rm S} &= 0 \,. \quad \hspace{0.45cm} (\text{temporal~gauge~condition}) \label{synchT}
\end{align}
In such a coordinate system, the fluid particle is at rest which amounts to use the comoving gauge conditions $v_{\rm C} = w^i_{\rm C}=0$.\footnote{In this paper, we do not explicitly use the comoving gauge condition but rather use it as a consistency check for the derivations in section~\ref{sec:PoissonToSynch}. Nevertheless, also since this is common in the literature, we shall call the gauge corresponding to the conditions~(\ref{synchS})--(\ref{synchT}) the ``synchronous-comoving gauge''.}
We note that the simultaneous conditions ``synchronous'' and ``comoving'' hold only for an irrotational and pressure-less fluid, and when neglecting any residual gauge modes \cite{malik&wands,bert96,Yoo:2014vta}. 

From the quote in the introduction one can deduce a general recipe to find any Eulerian gauge. A major subset of Eulerian gauges 
are defined with the spatial gauge \mbox{condition\,\cite{Rampf:2014mga}}
\be \label{generalEuler}
    E = F^i = 0 \,.
\ee
One reason for chosing this spatial gauge condition is the following. In the Lagrangian frame, the dynamical information of the displacement field is encoded in $E_{\rm S}$ and $F_{\rm S}^i$. 
Therefore, when performing a spatial gauge transformation to an Eulerian frame with vanishing $E$ and $F^i$ (see eq.\,\ref{generalEuler}), the spatial gauge generator must carry this dynamical information. 
In that very case, the spatial gauge generator is the displacement field.

Having specified only the spatial gauge condition is of course not sufficient for the study of general relativistic perturbations. A temporal gauge condition has to be imposed, which is, in principle arbitrary: Any temporal gauge condition supplemented with the
spatial gauge condition~(\ref{generalEuler}) fixes a unique Eulerian gauge.
Then, the resulting Eulerian equations of motion take, at least to first order, the Newtonian form in Eulerian coordinates \cite{Rampf:2013dxa}. 

In this paper we shall derive Eulerian solutions for two gauge choices, namely for the Poisson gauge and the total matter gauge. The Poisson gauge is defined by 
\begin{align}
    E_{\rm P} = F_{\rm P}^i  &= 0\,,  \qquad (\text{spatial~gauge~condition})  \\
    B_{\rm P} &=0  \,. \quad \hspace{0.45cm} (\text{temporal~gauge~condition})
\end{align}
It was originally introduced in ref.~\cite{bert96} as representing a generalisation of the so-called Newtonian (or longitudinal gauge) in order to account for vector and tensor modes, which are for example generated by the coupling of scalar modes beyond the linear regime.   
Note that the temporal condition corresponds to the minimal shear hypersurface condition in ref.\,\cite{Bardeen:1980kt} (see also \cite{bert96}).

The total matter gauge is another Eulerian gauge which is Eulerian in terms of the spatial coordinates, i.e., 
the spatial gauge conditions are given by $E_{\rm T} = F_{\rm T}^i = 0 $.
The temporal gauge condition is defined by requiring the vanishing of the total momentum potential on spatial hypersurfaces (see, e.g., \cite{malik&wands,Biern:2014zja}), i.e., the vanishing of the scalar part of the ${{}^0}_i$ component of the stress-energy tensor $T^\mu{}_\nu$. 
At first order, the perturbation of the total momentum is $T^0_{1i}= \left(\bar{\rho}+\bar{P}\right)\left(v_{1i}+B_{1i}\right)$, where $\bar{P}$ is the background pressure, and the resulting temporal gauge condition reads  
\begin{equation}
v_{1_{\rm T}}+B_{1_{\rm T}}=0\,.\label{T0i1}
\end{equation} 
At second order, the total momentum is \cite{malik&wands}
\begin{align}
T^0_{2i} &=
\left(\bar{\rho}+\bar{P}\right)\Big[v_{2i}+B_{2i}+
4C_{1ik}v_1^{~k}-2\psi_1\left(v_{1i}+2B_{1i}\right)\Big] \nonumber \\ &\qquad +2\left(\rho_1+
P_1\right)\left(v_{1i}+B_{1i}\right) +\frac{2}{a^2}\left(B^k_1 +
v_1^k \right)\Pi_{1ik} \,,
\end{align}
where  $\Pi_{ij}$ is the anisotropic stress tensor.
Assuming vanishing first-order vector modes and the spatial gauge condition of the total matter gauge, the requirement of the vanishing of the scalar part gives at second order
\begin{equation}
v_{2_{\rm T}} + B_{2_{\rm T}} -2 \nab^{-2} \partial^l \left( \psi_{1_{\rm T}} B_{1_{\rm T},l} + 2 \phi_{1_{\rm T}} v_{1_{\rm T},l}\right) = 0 \,. \label{T0i2}
\end{equation}
This temporal gauge condition is valid for a  (multi-component) fluid with pressure and non-vanishing anisotropic stress. Furthermore, we note that there exists an alternative definition 
for the temporal gauge condition of the total matter gauge \cite{Yoo:2014vta,Uggla:2013kya,Uggla:2014hva}, that is employing the vanishing of
the scalar part of the spatial component of the covariant 4-velocity, $u_i = g_{i0} u^0 + g_{ij}u^j$. 
Demanding $u_i =0$ for its scalar part leads to the same condition~(\ref{T0i1}) at first order, and to condition~(\ref{T0i2}) at second order,\footnote{We thank John Wainwright and Claes Uggla for pointing this out.} and thus this temporal gauge condition is formally equivalent with demanding the vanishing of the scalar part of ${T^0}_{i}$.
In summary, we define the total matter gauge with
\begin{align}
  E_{\rm T} = F_{\rm T}^i &= 0   \,, \qquad (\text{spatial gauge condition})  \\
 {\cal S}^i\, {T^0}_{i_{\rm T} }  &= 0\,, \qquad  (\text{temporal gauge condition})  \label{tempTOM}
\end{align}
where we have defined the scalar mode extraction operator ${\cal S}^i := \nab^{-2} \partial^i$
and, again, condition~(\ref{tempTOM}) amounts to use~(\ref{T0i1}) at first order and~(\ref{T0i2}) at second order.\footnote{In an earlier version of this manuscript, we have used the temporal gauge condition $v^0=0$ instead of~(\ref{tempTOM}). 
We stress that for an irrotational and pressure-less fluid the results in the total matter gauge are unaffected by choosing that condition or the other, yet it is highly non-trivial to prove the equivalence of these conditions on general grounds. We leave this open issue for future work.}
As we shall see in section~\ref{synchro2x} when we calculate the perturbations in the total matter gauge for an irrotational dust fluid, the time coordinate in the total matter gauge is the proper time of the fluid particle, as it is the case in the Lagrangian gauge. In the language of gauge transformations, this also means that the temporal gauge generator from synchronous-comoving gauge to the total matter gauge is precisely zero. 
This choice of Eulerian frame is thus very convenient when relating GR results to Newtonian investigations, since the relativistic Eulerian-Lagrangian correspondence makes use of the identical time coordinate, i.e., the proper time. So to say, the problem of relating a GR description to a Newtonian world is reduced from a 4D problem to a 3D problem.
Such considerations are for example useful when generating GR initial conditions for Newtonian $N$-body simulations (see \cite{FidlerEtAll}).

%%%%%%%%%%%%%%%%%%%%%%%%%%%%%%%%%%%%%%%%%%%%%%%%%%%%%%%%%%%%%%%%%%%%%%%%%%%%%%%%%%%%%%%%%%%%%%%%%%%%
\section{Evolution of perturbations in the Poisson gauge} \label{Poisson}
%%%%%%%%%%%%%%%%%%%%%%%%%%%%%%%%%%%%%%%%%%%%%%%%%%%%%%%%%%%%%%%%%%%%%%%%%%%%%%%%%%%%%%%%%%%%%%%%%%%%%%%
The Poisson gauge can be perturbatively defined by $B_{(r)_{\rm P}}=0$ for the temporal condition and by $E_{(r)_{\rm P}}=F_{i(r)_{\rm P}}=0$ for the spatial condition.
For second-order perturbations in the Poisson gauge in the Einstein--de Sitter (EdS) background see e.g.\ \cite{MMB} and \cite{Boubekeur:2008kn}, and for a non-vanishing cosmological constant see e.g.~\cite{Tomita:2005et,BMR2006}.  
Here we start by reporting and simplifying the results of ref.~\cite{BMR2006} for the metric and the four-velocity, and of ref.~\cite{BMPR2010} for the density contrast. 
The new expressions obtained here will be the starting point for the gauge transformation in the following sections. 
Let us just recall the procedure to solve the Einstein equations for the perturbations, which is the same at first and second order, see Appendix A of \cite{BMR2006}. The trace part of the $ij$ components of Einstein equations gives the evolution of the scalar perturbation $\phi_{(r)\rm P}$, and the trace-free part gives the relation with the other scalar $\psi_{(r) \rm P}$. The vector mode $\omega_{2i_{\rm P}}$ is found from the $0i$ components of Einstein equations and the tensor modes $\pi_{2ij_{\rm P}}$ obey an evolution equation obtained from the trace-less $ij$ components, once the scalar and vector perturbations are determined. Finally, the $00$ component of the Einstein equations and the $0i$ components, together with the momentum conservation, provide the density and four-velocity perturbations, respectively.

Following \cite{BMR2006}, we write the initial conditions in terms of the curvature perturbation of the uniform density hypersurfaces, $\zeta$. This is gauge invariant 
quantity and remains constant on super-horizon scales after it has been generated at the primordial epoch \cite{BHMW14}, and therefore provides the initial conditions for cosmological perturbations re-entering the Hubble radius at later time, including all the necessary information about primordial non-Gaussianity. We expand $\zeta$ as 
$\zeta =  \zeta^{(1)} + (1/2)\,\zeta^{(2)} =\zeta^{(1)} + (a_{\rm nl} -1)(\zeta^{(1)})^2 + \cdots$, where the parameter $a_{\rm nl}$ encodes the local primordial non-Gaussianity for different inflationary scenarios. 
Initial conditions are fixed deep in the matter-dominated era when the relation between the curvature perturbation with the gravitational potential is given by $\zeta_1=-5\varphi_{1\rm in}/3$.\footnote{This relation is sometimes also given in terms of the
Bardeen potential $\Phi$. Its relation to the linear gravitational potential is $\Phi=-\varphi_{1\rm in}$.} Our initial conditions up to second order are then written as 
\begin{eqnarray}
\zeta_{{\rm in}}&=&\zeta_{1{\rm in}}+\frac{1}{2}\zeta_{2{\rm in}} =  -\frac{5}{3}\varphi_{1\rm in}+\frac{25}{9}\left(a_{\rm nl}-1\right)\varphi^2_{1\rm in} \,, \label{init}
\end{eqnarray}
where $\zeta_{1{\rm in}}$ and $\varphi_{1\rm in}$ are first-order Gaussian random fields. The same initial conditions are written in terms of the initial gravitational potential as $\varphi_{\rm in}=\varphi_{1{\rm in}}+f_{\rm nl}\,\varphi^2_{1{\rm in}}$, where we can easily read-off the relation of the non-linear parameters, 
\be 
  f_{\rm nl}=\frac 5 3  \left( a_{\rm nl}-1 \right) \,.
\ee

\subsection{FLRW background equations}\label{sec:0}

Let us briefly summarise the results of the evolution equations for the FLRW background.
The energy constraint, Raychaudhuri equation and continuity equation for matter are
respectively
\begin{align}
3{\mathcal H}^2  &=8 \pi G a^2\overline{\rho}+a^2\Lambda  \,, \label{energy} \\
3\dot{{\mathcal H}} +4\pi G \overline{\rho} -a^2\Lambda&=0 \,,  \label{Raychaudhuri}\\
\dot{\overline{\rho}}+3{\mathcal H} \overline{\rho} &=0  \label{contunuity} \,,
\end{align} 
where $\Lambda$ is the cosmological constant and $\bar \rho \sim a^{-3}$ is the background density of matter.
We will make frequent use of the following definitions
\be \label{omegam}
  \Omega_{\rm m} \equiv\frac{8 \pi G a^2}{3{\mathcal H}^2}\overline{\rho} = \frac{{\mathcal H}_0^2\Omega_{\rm m0}}{a{\mathcal H}^2} \,, \qquad \Omega_\Lambda\equiv\frac{a^2\Lambda}{3{\mathcal H}^2}= \frac{ a^2 {\mathcal H}_0^2 \Omega_{\Lambda0}}{{\mathcal H}^2}\,,
\ee
where the subscript 0 indicates an evaluation at present time $\eta_0$, and we have defined $\Omega_{\rm m0}=8 \pi G a_0^2\overline{\rho}_{0}/(3 {\mathcal H}_0^2)$.
By using these definitions the Friedmann equations \eqref{energy} and \eqref{Raychaudhuri} read
\begin{align}
 {\mathcal H}^2 &= {\mathcal H}_0^2\left(\Omega_{\rm m0} a^{-1}+\Omega_{\Lambda 0}a^2\right)  \,, \\
  \dot{{\mathcal H}} &=  {\cal H}^2- \frac{3}{2} \frac{{\cal H}_0^2\Omega_{\rm m0}}{a} \,. \label{Rayd2}
\end{align}
In this paper we will always assume a flat Universe with $\Omega_{\rm m} + \Omega_\Lambda =1$.

\subsection{First-order results}

\paragraph{Metric tensor}
 Since we assume vanishing vector and tensor perturbations at first order, in this section we only consider scalar perturbations. 
At first order, the trace-free part of the $ij$ components of the Einstein equations implies that the two scalar perturbations in the metric coincide, i.e., 
\begin{equation}
\psi_{1_{\rm P}}=\phi_{1_{\rm P}}=\varphi \label{varphi}\,,
\end{equation}
and the trace part of the $ij$ component gives the evolution equation
\begin{equation} \label{evophi1P}
\ddot{\varphi}+3{\cal H}\dot{\varphi}+a^2\Lambda\varphi=0\,.
\end{equation}
The solution, selecting only the growing mode, can be written as
\begin{equation}
\label{relphiphi_0}
\varphi(\fett{x}, \eta)= g(\eta)\, \varphi_0(\fett{x}) \, ,
\end{equation}
where $\varphi_0$ is the peculiar gravitational potential linearly extrapolated to the present time $\eta_0$\footnote{Here and in what follows quantities with the subscript $0$ are meant to be evaluated at the present time.} and 
the time coefficient $g$ has still to be determined. Plugging the solution \eqref{relphiphi_0} in the evolution equation \eqref{evophi1P} and introducing the new unknown variable ${\cal D}=a g$, we obtain the differential equation
\begin{equation} \label{eqforD+} %\label{timecoeff1}
\ddot{\cal D} +\mathcal{H}\dot{\cal D}-\frac{3}{2} \frac{\mathcal{H}_0^2\Omega_{{\rm m}0}}{a} {\cal D} =0\,.
\end{equation}
which is nothing but the known evolution equation for the first-order time coefficient of the Newtonian density contrast \cite{Bernardeau:2001qr}. In this paper we are only interested in the growing mode solutions of ${\cal D}$ (and of the other time coefficients, of course). 
The solution for the first-order scalars is thus given by
$\varphi(\fett{x}, \eta)= g(\eta)\, \varphi_0(\fett{x})$,
where $g={\cal D}/a$ is identified to be the
growth-suppression factor, and ${\cal D}$ is the growth factor, namely the linear growing-mode solution of eq.\,\eqref{eqforD+} for the Newtonian density contrast $\delta_{1}(\eta, \fett{x})={\cal D}(\eta)\delta_{1}(\fett{x}, \eta_0)$. We note that this
identification implies that the solution in eq.\,(\ref{relphiphi_0}) coincides with the first-order Newtonian cosmological potential.
The analytical expression for the fastest growing mode ${\cal D}$ in terms of the scale factor is given by \cite{Enqvist:2010ex,Rigopoulos:2014rqa} 
\be \label{growth}
   \mathcal{D}(a) = \frac 5 2 {\cal H}_0^2 \Omega_{{\rm m}0} \frac{{\cal H}}{a} \int_0^a \frac{\dd x}{{\cal H}^3(x)} = a \sqrt{1 + \frac{\Omega_{\Lambda0}}{\Omega_{{\rm m}0}} a^3} \,{}_2F_1 \left( \frac 3 2 , \frac 5 6, \frac{11}{6}, - \frac{\Omega_{\Lambda0}}{\Omega_{{\rm m}0}} a^3 \right) \,,
\ee
where ${}_2F_1$ is the Gauss hypergeometric function. 
We normalise the growth factor such that ${\cal D}(\eta_0)=1$.  We finally consider the Newtonian Poisson equation
\be  \label{PoissonEq}
\nab^2\varphi(\fett{x}, \eta)-\frac{3}{2} \frac{{\cal H}_0^2 \Omega_{\rm m0}}{a}\delta_1( \fett{x}, \eta) =0 \,,
\ee
evaluated at present time to fix the constant $\delta_{1}(\fett{x}, \eta_0)$, and we find
\begin{equation} \label{delta1N0}
\delta_{1}(\fett{x}, \eta_0)=\frac{2}{3}\frac{\nab^2\varphi_0(\fett{x})}{\mathcal{H}_0^2\Omega_{{\rm m}0}}\,.
\end{equation}
In an EdS universe we have  ${\cal D}(\eta)=\eta^2$ and $\stuff =4$, so $\delta_1 = (\eta^2/6) \nab^2 \varphi_0$.

\paragraph{Four-velocity} The spatial component of the peculiar four-velocity is given by
\begin{equation}
v_{1_{\rm P}}  = - \frac{1}{4 \pi G a^2 \bar\rho} 
\left(\dot{\varphi} + {\cal H} \varphi\right)  = - \frac{2}{3 {\cal H}_0^2 \Omega_{\rm m0}} \dot {\cal D} \varphi_{0} \,, \label{v1P} 
\end{equation}
which contains the scalar part only (the vector part of the velocity is vanishing at this order). 
This coincides with the first-order Newtonian peculiar velocity in
Newtonian Eulerian perturbation theory (NEPT).
To obtain the perturbation of the zeroth component of the four-velocity, we 
plug the result~(\ref{relphiphi_0}) in Eq.\,(\ref{v0any}) and immediately obtain
\be
  v^0_{1_{\rm P}} \equiv - \psi_{1_{\rm P}} = - g\, \varphi_0 \,.
\ee

\paragraph{Matter density} The final expression for the density contrast is
\begin{equation}
\delta_{1_{\rm P}} = \frac{2}{3 {\cal H}_0^2 \Omega_{\rm m0}} \left[ {\cal D} \nab^2 \varphi_0 - 3 {\cal H} \dot {\cal D} \varphi_0 \right]\, .
\end{equation}
The first term in the above equation is identical to the first-order Newtonian
density contrast, whereas the second term is a GR correction.
Actually, as we shall see in section~\ref{FO_PtoS}, this GR correction is 
the result of the different time coordinate used in the Poisson gauge, which is due to the relative velocity between this Eulerian frame 
and the Lagrangian frame, see in particular eq.\,\eqref{rho1} and eq.\,\eqref{B1s}.

\subsection{Second-order results}
\paragraph{Metric tensor: scalar perturbations} Our starting point are the results of \cite{BMR2006} and \cite{BMPR2010}. However, as we shall see, it is possible to simplify their expressions tremendously. We begin with eqs.~(2.22) and~(2.23) of \cite{BMR2006}\begin{eqnarray}
\label{psi2p}
\psi_{2_{\rm P}}(\eta, \fett{x})&=&\left( {\cal B}_1+4g^2
-2g g_{\rm in} -\frac{10}{3}(a_{\rm nl}-1)g g_{\rm in}\right)\varphi_0^2 +6\Bigg[ {\cal B}_2 +\frac{4}{3} g^2  \left( e+\frac{3}{2} \right) \nonumber \\
&-&\frac{4}{3}g g_{\rm in} \Bigg] \Theta_{0} 
+{\cal B}_3 \nab^{-2} \partial_i\partial^j(\partial^i \varphi_0 \partial_j \varphi_0 )+{\cal B}_4 \partial^i \varphi_0 \partial _i\varphi_0\, ,\\
\label{phi2p}
\phi_{2_{\rm P}}(\eta,\fett{x})&=&\left( 
{\cal B}_1-2g g_{\rm in} -\frac{10}{3}(a_{\rm nl}-1)g g_{\rm in}
\right)\varphi_0^2 
+6\left( {\cal B}_2 -\frac{4}{3}g g_{\rm in}  \right) \Theta_{0} \nonumber \\
&+&{\cal B}_3  \nab^{-2} \partial_i\partial^j(\partial^i \varphi_0 \partial_j 
\varphi_0 )+{\cal B}_4 \partial^i \varphi_0 \partial _i\varphi_0 \,,
\end{eqnarray}
where the subscript ``in'' denotes an evaluation at initial time (given deep in the matter dominated era), $e \equiv f^2/\Omega_{\rm m}$ with $f \equiv \dd \ln {\cal D} / (\dd \ln a)$, and 
\begin{equation}
\Theta_{0}= \frac{1}{2} \nab^{-2}  \Bigg[ \frac 1 3 \varphi_0^{,l}  \varphi_{0,l} -  \nab^{-2} \left( \varphi_0^{,l}  \varphi_{0}^{,m} \right)_{,lm} \Bigg] \,. \label{Theta0}
\end{equation}
The time coefficients ${\cal B}_i(\eta)$ are reported in \cite{BMR2006} in eqs.\,(2.24) and (2.25) (in the notation of \cite{BMR2006}, these coefficients are $B_i$). They involve unfortunately very complicated integrals over time with no explicit solutions given.
Here we find explicit solutions for these time coefficients. Let us begin with the PDE for the scalar perturbation $\phi_{2_{\rm P}}$ which is the trace of the $ij$ components of the Einstein equations, eq.\,(2.7) in \cite{BMR2006}
\be \label{eq:ODE}
  \ddot{\phi}_{2_{\rm P}}+3{\cal H}\dot{\phi}_{2_{\rm P}}+a^2\Lambda\phi_{2_{\rm P}}=S(\eta, \fett{x}) \,,
\ee
with the source term
\begin{align}
\label{PSIsource}
S(\eta,\fett{x})=g^2\Omega_{\rm m} {\mathcal H}^2 \Bigg[\frac{(f-1)^2}{\Omega_{\rm m}} 
\varphi_0^2+12 \Bigg(2 \frac{(f-1)^2}{\Omega_{\rm m}} -\frac{3}{\Omega_{\rm m}}+3 \Bigg) \Theta_0\Bigg]& \nonumber \\
+ g^2 \Bigg[ \frac{4}{3}   
\left( \frac{f^2}{\Omega_{\rm m}}+\frac{3}{2} \right) \nab^{-2}\partial_i\partial^j\left(\partial_j\varphi_0\partial^i\varphi_0\right)- \partial_i\varphi_0\partial^i\varphi_0 \Bigg]& \, .
\end{align}
To obtain the ODE's for the ${\cal B}_i$'s we impose the Ansatz 
\be \label{Ansatz}
  \phi_{2_{\rm P}}(\eta, \fett{x})  = {\cal B}_1(\eta) \,\phi_{2_{\rm P}}^{(1)}(\fett{x}) +
   {\cal B}_2(\eta) \,\phi_{2_{\rm P}}^{(2)}(\fett{x}) +  {\cal B}_3(\eta) \,\phi_{2_{\rm P}}^{(3)}(\fett{x}) 
   + {\cal B}_4(\eta) \,\phi_{2_{\rm P}}^{(4)}(\fett{x})+\frac{g(\eta)}{g_{\rm in}} \phi_{2_{\rm P_{\rm in}}}(\fett{x})\,,
\ee
where
\begin{align}
 \begin{split}
  \phi_{2_{\rm P}}^{(1)}(\fett{x}) &= \varphi_0^2 \,, \qquad \hspace{2.87cm}
  \phi_{2_{\rm P}}^{(2)}(\fett{x}) = \Theta_0  \,,  \\
  \phi_{2_{\rm P}}^{(3)}(\fett{x}) &=   \nab^{-2} \partial_i\partial^j(\partial^i \varphi_0 \partial_j  \varphi_0 ) \,, \qquad 
  \phi_{2_{\rm P}}^{(4)}(\fett{x}) = \partial^i \varphi_0 \partial _i\varphi_0 \,,
 \end{split}
\end{align}
and the last term is the solution of the homogeneous equation with $\phi_{2_{\rm P_{\rm in}}}(\fett{x})$ representing the initial condition taken deep in matter-dominated era and on super-horizon scales. We now repeat the same procedure we used to determine the time coefficient of the first-order scalar $\phi_{1_{\rm P}}$, given 
in eq.\,\eqref{relphiphi_0}.
For demonstrative purposes, we briefly show how to solve for the ${\cal B}_1$. We plug ${\cal B}_1 \phi_{2_{\rm P}}$ into~(\ref{eq:ODE}) and keep only terms proportional to $\phi_{2_{\rm P}}^{(1)}= \varphi_0^2$. The temporal and spatial dependence thus factorise so that the
 ODE for ${\cal B}_1$ is
\be
\ddot{{\cal B}}_1+3{\cal H}\dot{{\cal B}}_1+a^2\Lambda {\cal B}_1=g^2\mathcal{H}^2(f-1)^2 \,,
\ee
or, introducing the new unknown variable $b_1\equiv a {\cal B}_1$, equivalently
\begin{align}
  \ddot b_1 + {\cal H} \dot b_1  - \frac 3 2 \frac{ {\cal H}_0^2 \Omega_{\rm m0}}{a} b_1  &= \frac{{\cal D}^2 {\cal H}^2}{a} (f-1)^2 \,, \label{PDEb1}
\intertext{where we have used eq.\,(\ref{Rayd2}).
Likewisely, we obtain for the other $b_i\equiv a {\cal B}_i$ }
   \ddot b_2 + {\cal H} \dot b_2  - \frac 3 2 \frac{ {\cal H}_0^2 \Omega_{\rm m0}}{a} b_2  &= \frac{2 {\cal D}^2 {\cal H}^2}{a} \left[ 2(f-1)^2 + 3( \Omega_{\rm m}-1) \right] \,, \\
   \ddot b_3 + {\cal H} \dot b_3  - \frac 3 2 \frac{ {\cal H}_0^2 \Omega_{\rm m0}}{a} b_3 &= 2 \frac{{\cal D}^2}{a} + \frac{4 \dot {\cal D}^2}{3{\cal H}_0^2 \Omega_{\rm m0} } \,, \label{eq:b3} \\
  \ddot b_4 + {\cal H} \dot b_4  - \frac 3 2 \frac{ {\cal H}_0^2 \Omega_{\rm m0}}{a} b_4 &= - \frac{{\cal D}^2}{a} \,. \label{PDEb4}
\end{align}
To arrive at eq.\,(\ref{eq:b3}) we have used the ODE for the first-order growth ${\cal D}$, eq.\,(\ref{eqforD+}).
The easiest solution is obtained for $b_4$, by noting that its ODE can be written as
\be \label{eqF}
 \ddot{\cal F} +\mathcal{H}\dot{\cal F}-\frac{3}{2} \frac{\mathcal{H}_0^2\Omega_{\rm m0}}{a} {\cal F }=\frac{3}{2}  \frac{\mathcal{H}_0^2\Omega_{\rm m0}}{a} {\cal D}^2 \,,
\ee
with ${\cal F} = - (3/2) \mathcal{H}_0^2\Omega_{\rm m0} b_4$. Equation~(\ref{eqF}) is nothing but the known evolution equation for the second-order time coefficient of the Newtonian displacement field.\footnote{\label{footdisplacement}
Note explicitly, that ${\cal F}$ is not the second-order growth of the density because the density does not factorise into a single time and space-dependent part. By contrast, the displacement field does:
In the notation of eq.\,(\ref{NLPT}), the Newtonian displacement field is up to second
order $\bar{\Psi}_i= {\cal D}(\bar\eta) \bar{\Psi}_{1i}(\bar{\fett{q}})
 + {\cal F}(\bar\eta) \bar{\Psi}_{2i}(\bar{\fett{q}})$, where ${\cal D}$ and ${\cal F}$ are truly the
solutions of eqs.\,(\ref{eqforD+}) and~(\ref{eqF}), respectively. For the Newtonian
trajectory see eq.\,(\ref{NLPTfinal}).
}
Its solution for the fastest growing mode in a $\Lambda$CDM Universe can be found in  \cite{Matsubara:1995kq}. For an EdS universe we have ${\cal F}=(3/7)\eta^4$.
The solutions for $b_1$, $b_2$, and $b_3$ are also easily obtained. We find
the fairly simple growing mode solutions
\begin{align}
 b_1 &= -{\cal D}g + \frac{2\dot {\cal D}^2}{3{\cal H}_0^2 \Omega_{\rm m0}} + \frac 1 3 {\cal D} g_{\rm in}  \,,  \label{solb1}\\
 b_2 &= - 2 {\cal D}(g_{\rm in} - g ) \,, \label{solb2} \\
 b_3 &= \frac{2}{3{\cal H}_0^2 \Omega_{\rm m0}} \left( {\cal F}+ {\cal D}^2 \right) \,, \\
 b_4 &= - \frac{2}{3{\cal H}_0^2 \Omega_{\rm m0}} {\cal F} \label{solb4} \,.
\end{align}
Note that for an EdS universe, $b_1=b_2=0$, whereas $b_3 =(5/21)\eta^4$ and $b_4=-(1/14) \eta^4$.

With the above simplifications, we can finally write the two scalar perturbations as
\begin{align}
  \psi_{2_{\rm P}} &= \left( 3 g^2 + \frac 5 3 g g_{\rm in} (1- 2a_{\rm nl})  + \frac{2 \dot {\cal D}^2}{3a \mathcal{H}_0^2\Omega_{\rm m0}}  \right)\varphi_0^2 %\nonumber \\ &
+ 6 \left(4 g^2 - \frac{10}{3} g g_{\rm in} + \frac 4 3 \frac{\dot {\cal D}^2}{a \mathcal{H}_0^2\Omega_{\rm m0}} \right) \Theta_0 \nonumber \\
 &\quad\hspace{2cm}+ \frac{2{\cal D}^2}{3a\mathcal{H}_0^2\Omega_{\rm m0}}  \varphi_{0,l} \varphi_0^{,l}-\frac{4\left({\cal D}^2+{\cal F}\right)}{3a\mathcal{H}_0^2\Omega_{\rm m0}}\Psi_0 \,, \label{eq:psi2P}\\
 \phi_{2_{\rm P}} &=  \left( - g^2 + \frac 5 3 g g_{\rm in} (1- 2a_{\rm nl})  + \frac{2 \dot {\cal D}^2}{3a \mathcal{H}_0^2\Omega_{\rm m0}}  \right)\varphi_0^2 %\nonumber \\&
+ 6\left( 2 g^2 -\frac{10}{3}g g_{\rm in}  \right) \Theta_{0} \nonumber \\
 &\quad\hspace{2cm}+\frac{2{\cal D}^2}{3a\mathcal{H}_0^2\Omega_{\rm m0}} \varphi_{0,l} \varphi_0^{,l}-\frac{4\left({\cal D}^2+{\cal F}\right)}{3a\mathcal{H}_0^2\Omega_{\rm m0}}\Psi_0  \label{eq:phi2P}\,,
\end{align}
where 
\begin{align}
 \Psi_0 &= - \nab^{-2}\frac{1}{2} \left[(\nab^2\varphi_0)^2- \varphi_{0,ik}   \varphi_0^{,ik} \right] \,.  \label{Psi0}
\end{align}
The last line in~(\ref{eq:psi2P}) and~(\ref{eq:phi2P}) is the second-order gravitational potential in NEPT (see, e.g., \cite{Matsubara:1995kq}), whereas the remnant terms are relativistic corrections.

Note that in deriving the last line of~(\ref{eq:psi2P}) (and the last line of~(\ref{eq:phi2P})), 
we have made use of the identity
\be
  \nab^{-2} \partial_l\partial^m(\partial^l \varphi_0 \partial_m  \varphi_0 )
   = -2 \Psi_0 + \varphi_0^{,l} \varphi_{0,l} \,.  
\ee
Using this identity in~(\ref{eq:phi2P}) we obtain the relation between the ``GR kernel`` 
$\Theta_0$ and the ``Newtonian kernel'' $\Psi_0$:
\be
  \nab^2 \Theta_0 = \Psi_0 - \frac 1 3 \varphi_{0,l} \varphi_0^{,l} \,, \label{relationTheta0}
\ee
which will be helpful in the following sections.

For the limiting case of an EdS universe with $a_{\rm nl}=0$, our results~(\ref{eq:psi2P}) and~(\ref{eq:phi2P}) agree with \cite{MMB}. For $\Lambda$CDM, the
expressions~(\ref{eq:psi2P}) and~(\ref{eq:phi2P}) have been first derived in refs.~\cite{Uggla:2014hva,Uggla:2013kya}: Although different spatial and temporal functions\footnote{In particular, their temporal function ${\cal B}$ is related to our ${\cal F}$ and ${\cal D}$ via ${\cal B} = (1/3) [1- 2{\cal F}/{\cal D}^2]$. We thank John Wainwright and Claes Uggla for pointing this out.} are used, our result agrees with theirs.

\paragraph{Metric tensor: vector perturbations}
In a $\Lambda$CDM Universe, linear vector perturbations are usually discarded. But even in the absence of linear vectors, the non-linear gravitational evolution generates second-order vector perturbations, as it is well-known. These second-order vector perturbations arise from quadratic couplings of first-order scalar perturbations.
For an EdS universe, the first derivation of these vector perturbations was given in \cite{MMB}, whereas the results for a $\Lambda$CDM Universe were reported in \cite{BMR2006}. To get these vector perturbations, one possible derivation is to
first observe that the second-order momentum conservation equation ($T^\mu_{i;\mu} =0$, see e.g., eq.\,(A.26) in \cite{BMR2006}) is only sourced by a pure scalar for growing mode initial conditions. From that one can easily deduce that the vector perturbation $\omega_{2i_{\rm P}}$ has to be equal minus the transverse part of the velocity, $-w_{2i_{\rm P}}$. Then, $\omega_{2i_{\rm P}}$ can be extracted from the $0i$ component of the Einstein equations which is \cite{BMR2006}\footnote{Note that in deriving this expression we have corrected two typos in eq.\,(A.27) of \cite{BMR2006}.}
 \be
  {\cal H} \psi_{2_{\rm P},i} + \dot \phi_{2_{\rm P},i} - \frac 1 4 \nab^2 \omega_{2i_{\rm P}}
    + 10 g \dot g \varphi_0 \varphi_{0,i} = - 3  {\cal H}^2  \Omega_{\rm m} \left[  (g \varphi_0 + \delta_{1_{\rm P}}) v_{1_{\rm P},i} + \frac{v_{2_{\rm P},i}}{2} \right] \,.  \label{A27}
\ee

Extracting from this expression the transverse part we then obtain
\begin{equation} \label{omega2ip}
 \omega_{2i_{\rm P}}=- \frac{16}{3{\cal H}_0^2\Omega_{\rm m0}} \frac{{\cal D} \dot {\cal D}}{a} {\cal R}_i (\fett{x})\,,
\end{equation}
which is purely relativistic, and we have defined the transverse kernel ($\nab \cdot \fett{\cal R}=0$)
\be \label{transR}
  {\cal R}_i(\fett{x}) =  \nab^{-2}\left( \varphi_{0,i} \nab^2\varphi_0 - \varphi_0^{,l} \varphi_{0,li} + 
2 \Psi_{0,i} \right) \,.
\ee

\paragraph{Metric tensor: tensor perturbations}
 
In the Poisson gauge, the tensor perturbations obey the wave equation \cite{BMR2006}
\be \label{waveeq}
  \ddot \pi_{2ij_{\rm P}} + 2 {\cal H} \dot \pi_{2ij_{\rm P}}  - \nab^2 \pi_{2ij_{\rm P}}  = s(\eta) \, {\cal T}_{ij}(\fett{x}) \,,
\ee
where $s(\eta) = -8 g^2 (1+2f^2/(3 \Omega_{\rm m}))$,  and ${\cal T}_{ij}(\fett{x}) \equiv \nab^{-2} {\cal S}_{ij}$ is a divergenceless and traceless tensor related to 
\be \label{tensorSij}
  {\cal S}_{ij}(\fett{x}) = \nab^2 \Psi_0 \delta_{ij} +\Psi_{0,ij} +2 (\varphi_{0,ij}\nab^2 \varphi_0 - \varphi_{0,ik}\varphi_{0,j}^{,k} ) \,.
\ee
The solution of~(\ref{waveeq}) is  
\be \label{GW}
  \pi_{2 ij_{\rm P}}(\eta,\fett{x}) = \int \frac{\dd^3 \fett{k}}{(2\pi)^3} \tilde \pi_{2 ij_{\rm P}}(\eta,\fett{k})\, {\rm e}^{{\rm i} \fett{k} \cdot \fett{x}} \,,
\ee
with
\be 
  \tilde \pi_{2 ij_{\rm P}}(\eta,\fett{k}) = \tilde {\cal T}_{ij} (\fett{k}) \, \tilde h(\eta,k) \,,
\ee
where $k=|\fett{k}|$,  and\footnote{Note that we have corrected a typo in the last integrand of eq.\,(4.11) in \cite{BMR2006}.}
\begin{align}
 \tilde  h(\eta,k) &= \tilde\chi_1(\eta,k) \int \dd \tilde \eta \frac{\tilde \chi_2(\tilde\eta,k)}{W} s(\tilde \eta) -  \tilde\chi_2(\eta,k) \int \dd \tilde \eta \frac{\tilde \chi_1 (\tilde\eta,k)}{W} s(\tilde \eta)  \,,
\end{align}
where $\tilde\chi_1$ and $\tilde\chi_2$ are the solutions of 
\be
  \ddot {\tilde\chi} + 2 {\cal H} \dot {\tilde\chi} + k^2 \tilde \chi =0 \,,
\ee
with corresponding Wronskian $W = \dot{\tilde\chi}_1 \tilde \chi_2 - \tilde \chi_1 \dot{\tilde\chi}_2$.

\paragraph{Four-velocity}
To our knowledge there are no explicit solutions in the literature for the second-order velocity in the Poisson gauge for the $\Lambda$CDM Universe. We take the divergence and the curl from eq.\,(\ref{A27}), then we obtain respectively  
\begin{align}
    v_{2_{\rm P}} &=  - \frac{2a}{3 \mathcal{H}_0^2\Omega_{\rm m0}}  \left(  {\cal H} \psi_{2_{\rm P}} + \dot \phi_{2_{\rm P}}   + 5 g \dot g \varphi_0^2 \right)  + \frac{2 g\dot {\cal D}}{3\stuff} \varphi_0^2 - \frac{4 {\cal H} \dot {\cal D}^2}{3(\stuff)^2} \varphi_0^2 \nonumber \\ &\quad
  - \frac{16 {\cal D} \dot {\cal D}}{9 (\stuff)^2} \Psi_0 + \frac{4 {\cal D} \dot {\cal D}}{9(\stuff)^2} \varphi_0^{,l} \varphi_{0,l} \nonumber \\
 &=  \frac{2 \dot {\cal D}}{3\stuff} \left[ \frac{10}{3} g_{\rm in} (a_{\rm nl} -1)  -3g \right] \varphi_0^2  - \frac{8 \dot {\cal D} g}{\stuff} \Theta_0 \nonumber \\
& \quad - \frac 4 9 \frac{{\cal D} \dot {\cal D}}{( \mathcal{H}_0^2\Omega_{\rm m0})^2} \varphi_{0,l} \varphi_{0}^{,l}
 + \frac 8 9 \frac{\dot {\cal F}}{( \mathcal{H}_0^2\Omega_{\rm m0})^2} \Psi_0 \,,  \label{eqv2p}
 \end{align}
and
\begin{equation}
w_{2i_{\rm P}} = -\omega_{2i_{\rm P}}= \frac{16}{3{\cal H}_0^2\Omega_{\rm m0}} \frac{{\cal D} \dot {\cal D}}{a}  {\cal R}_i \,,
\end{equation}
where we remind the reader that the transverse kernel ${\cal R}_i(\fett{x})$ is defined
in eq.\,(\ref{transR}). Note that the last line in eq.\,(\ref{eqv2p}) is the second-order peculiar velocity in NEPT (see, e.g., \cite{Matsubara:1995kq}).
To arrive at the final expression for $v_{2_{\rm P}}$, we have made use of the identity
\be \label{Halpha}
  \frac{{\cal H} \dot {\cal D}}{\mathcal{H}_0^2\Omega_{\rm m0}} = \frac 5 2 g_{\rm in} - \frac 3 2 g \,,
\ee
which we shall prove later, see in particular the derivation of eq.\,(\ref{eq:constIN}). 
For the perturbation of the zeroth component of the four-velocity, we 
plug the result~(\ref{eq:psi2P}) in Eq.\,(\ref{v0any}) and obtain 
\begin{align}
  v^0_{2_{\rm P}} &= - \left( \frac 5 3 g g_{\rm in} \left( 1 - 2 a_{\rm nl} \right) + \frac{2 \dot {\cal D}^2}{3a \stuff} \right)  \varphi_0^2  - 6 \left( 4 g^2 - \frac{10}{3} g g_{\rm in} + \frac{4 \dot {\cal D}^2}{3a \stuff} \right) \Theta_0 \nonumber \\
  &\quad+ \left(  \frac{4 \dot {\cal D}^2}{9(\stuff)^2} - \frac{2{\cal D}^2}{3 a \stuff} \right) \varphi_{0,l} \varphi_0^{,l} + \frac{4\left( {\cal D}^2 + {\cal F}\right)}{3a \stuff} \Psi_0
 \,.
\end{align}

We have checked that the EdS limit of the above results agree with the ones in \cite{MMB,BMR2005}.

\paragraph{Matter density}
With the explicit solutions of the time coefficients $b_i$ given in eqs.\,(\ref{solb1})--(\ref{solb4}) we are able to simplify the result given in ref.~\cite{BMPR2010}\footnote{Note that the first and the last coefficients of $\varphi^2=g^2\varphi_0^2$ in the first line of ref.~\cite{BMPR2010} eq.\,(29) are wrong. They have to be replaced by $2 (f-1)^2$. } to the fairly compact expression
\begin{align}
\label{rho2pNew}
\delta_{2_{\rm P}}&=  \frac{2{\cal H}^2 {\cal D}^2}{a {\cal H}_0^2 \Omega_{\rm m0}} \left[ f^2 -4f \right] \varphi_0^2 + \frac{10 {\cal H}}{3{\cal H}_0^2 \Omega_{\rm m0}} (1+2a_{\rm nl}) \dot {\cal D} g_{\rm in} \varphi_0^2 -\frac{24 {\cal H} \dot {\cal D} {\cal D}}{a {\cal H}_0^2 \Omega_{\rm m0}} \Theta_0  \nonumber\\
&+ \frac{2{\cal D}}{{\cal H}_0^2 \Omega_{\rm m0}} \left[ g - \frac{20}{9} g_{\rm in} a_{\rm nl}  \right] (\nab \varphi_0)^2 +\frac{8 {\cal H} \dot {\cal F}}{3({\cal H}_0^2 \Omega_{\rm m0})^2} \Psi_0\nonumber \\
&+  \frac{2}{3{\cal H}_0^2 \Omega_{\rm m0}}\left[ 6 {\cal D} g +\frac{10}{3} {\cal D} g_{\rm in} (1- 2a_{\rm nl}) + \frac 4 3 \frac{\dot {\cal D}^2}{{\cal H}_0^2 \Omega_{\rm m0}}  \right]  \varphi_0 \nab^2 \varphi_0 \nonumber \\
 &+ \frac{4}{9 ({\cal H}_0^2 \Omega_{\rm m0})^2} \Bigg[ ({\cal D}^2+{\cal F}) (\nab^2\varphi_{0})^2 + 2 {\cal D}^2 \varphi_{0}^{,l} \nab^2 \varphi_{0,l} + ({\cal D}^2 - {\cal F}) \varphi_{0,lm} \varphi_0^{,lm} \Bigg] \,.
\end{align}
We have checked that the EdS limit of the above agrees with the results of refs.\,\cite{MMB,BMR2005}. For $\Lambda$CDM, our result is in agreement with the ones reported in \cite{Uggla:2014hva,Uggla:2013kya}, however care has to be taken when comparing our results to theirs, because of different notations and normalisations used; for a clarification of this we refer the reader to section 4.4.3 in \cite{Uggla:2014hva}.
Note that the last line in eq.\,(\ref{rho2pNew}) is the second-order density contrast in NEPT (see, e.g., \cite{Matsubara:1995kq}).

Finally let us conclude this section with a comment about our solution scheme for Eq.\,(\ref{eq:ODE}) and its applicability to cosmological models beyond $\Lambda$CDM. 
The general solution technique for $\phi_{2_{\rm P}}$ (which comes with the Ansatz~(\ref{Ansatz})) should also be fruitful for other cosmological models, as long as the source term $S$ of the PDE for $\phi_{2_{\rm P}}$, Eq.\,(\ref{eq:ODE}), can be written in factorisable form. 
Note however that the source term cannot be factorised in e.g. multi-fluid models, or when a radiation component is included in the analysis, because then the respective growth functions are generally scale-dependent (see e.g.~\cite{Lesgourgues:2012uu} for the massive neutrinos case).

%%%%%%%%%%%%%%%%%%%%%%%%%%%%%%%%%%%%%%%%%%%%%%%%%%%%%%%%%%%%%%%%%%%%%%%%%%%%%%%%%%%%%%%
\section{From the Poisson gauge to the synchronous-comoving gauge}\label{sec:PoissonToSynch}
%%%%%%%%%%%%%%%%%%%%%%%%%%%%%%%%%%%%%%%%%%%%%%%%%%%%%%%%%%%%%%%%%%%%%%%%%%%%%%%%%%%%%%

In this section we apply the general formulas of section \ref{gaugetformulas} to the transformation from the Poisson gauge to the synchronous-comoving gauge, which is defined by $\psi_{(r)_{\rm S}}=0$ for the temporal condition and by $B_{(r)_{\rm S}}=\omega_{i(r)_{\rm S}}=0$ for the spatial condition. We define the synchronous-comoving gauge as the Lagrangian frame of reference of the fluid particle. The respective Lagrangian coordinates are  denoted with $(\tau,\fett{q})$.

Let us note that the equations for the metric transformed from any gauge to the synchronous-comoving gauge have always the following structure: at any fixed order the equations for the scalar in $g_{00}$ and for the scalar and the vector in $g_{0i}$ are four equations for the unknowns $\alpha_{(r)}$, $\beta_{(r)}$ and $d^i_{(r)}$, i.e., they give the functions of the transformation $\xi^\mu_{(r)}$. These results then give the spatial metric in the synchronous-comoving gauge. The equations for $\alpha_{(r)}$, $\beta_{(r)}$ and $d^i_{(r)}$ contain time derivatives, and the integration constants in the results represent the residual gauge freedom of the synchronous-comoving gauge. It can be set to zero by requiring that at initial time spatial and temporal coordinates coincide in the Lagrangian and Eulerian gauges, i.e., the Lagrangian and Eulerian positions coincide initially. We shall make frequent use of this requirement. 

Before beginning with the calculations, let us emphasise that our results
for the gauge generators and for the metric in the synchronous-comoving gauge are entirely new and they do agree with the findings in the literature in the restricted case of an EdS universe, \cite{Bartolo:2005xa,MMB}.
In addition, by transforming our result for the density contrast in the Poisson gauge, eq.\,(\ref{rho2pNew}), we are able to derive a fairly simple result for the density contrast in the synchronous-comoving gauge.  
Note especially that we are able to solve for all time coefficients which
actually coincide
with the first-order and second-order structure growth of the Newtonian displacement field, respectively denoted with ${\cal D}$ and ${\cal F}$ (see footnote~\ref{footdisplacement} on page~\pageref{footdisplacement}).

\subsection{First-order transformations}\label{FO_PtoS}
\paragraph{Metric tensor}
From eqs. \eqref{psi1}--\eqref{E1} we find
\bea
0&=& \psi_{1_{\rm P}}+\mathcal{H}\alpha_{1_{\PS}}+\dot{\alpha}_{1_{\PS}} \,, \label{psi1s}\\
\phi_{1_{\rm S}}&=&\phi_{1_{\rm P}}-\mathcal{H}\alpha_{1_{\PS}} -\frac{1}{3}\nab^2 \beta_{1_{\PS}}  \,, \label{phi1s}\\
\alpha_{1_{\PS}} &=&\dot{\beta}_{1_{\PS}}  \,, \label{B1s}\\
E_{1_{\rm S}}&=&\beta_{1_{\PS}}  \,. \label{E1s}
\eea
From \eqref{psi1s} and \eqref{B1s} we find the solution for $\alpha_{1_{\PS}}$ and $\beta_{1_{\PS}}$
\bea
\alpha_{1_{\PS}} &=& -\left(\frac{1}{a}\int_{\eta_{\rm in}}^\eta \dd\tilde{\eta} \,{\cal D}\right) \varphi_0 +I_1 \,,\\
\beta_{1_{\PS}} &=& -\int_{\eta_{\rm in}}^\eta \dd\eta'\, \left(\frac{1}{a}\int_{\eta_{\rm in}}^{\eta'} \dd\tilde{\eta} \,{\cal D}\right) \varphi_0 + I_2 \,,
\eea
where the space-dependent integration constants $I_1$ and $I_2$ represent the well-known residual gauge freedom of the synchronous-comoving gauge which is fixed from the initial conditions, by demanding that spatial and temporal coordinates coincide at initial time. Thus $I_1=I_2=0$. Using the identity
\begin{equation}
\dot{\cal D}=\frac{3}{2} \frac{{\cal H}_0^2\Omega_{{ \rm m}0}}{a}\int_{\eta_{\rm in}}^\eta \dd\tilde{\eta} \,{\cal D}(\tilde{\eta} )  \,, \label{magic4D}
\end{equation}
which is just the first integral of the equation \eqref{eqforD+} for the growth factor, we get
\bea
\alpha_{1_{\PS}} &=& %-\dot{D} (\tau)\nab^{-2}C(\fett{q})=
    -\dfrac{2}{3} \dfrac{\dot{\cal D}\varphi_0}{\mathcal{H}_0^2\Omega_{\rm m0}} \,, \label{ouralpha1s}\\
\beta_{1_{\PS}} &=& %-D(\tau)\nab^{-2}C(\fett{q})=
   -\dfrac{2}{3} \dfrac{{\cal D}\varphi_0}{\mathcal{H}_0^2\Omega_{\rm m0}}\label{ourbeta1s}\,.
\eea
Finally, we find $\phi_{1_{\rm S}}$ and $E_{1_{\rm S}}$ by substituting the last two expressions in \eqref{phi1s} and \eqref{E1s}
\bea
\phi_{1_{\rm S}} &=& \frac 2 9 \frac{{\cal D} \nab^2 \varphi_0}{\stuff}+g\varphi_0-{\cal H}\alpha_{1_{\PS}}  \,, \\ 
E_{1_{\rm S}} &=& %-\nab^{-2}C(\fett{q})D(\tau) %=-\nab^{-2}\delta_1(\tau, \fett{q}) =
-\dfrac{2}{3} \dfrac{{\cal D}\varphi_0}{\mathcal{H}_0^2\Omega_{\rm m0}} \,. \label{ourE1s}
\eea
In our expression for $\phi_{1_{\rm S}}$ the quantity
\begin{equation} \label{eq:const}
g\varphi_0-{\cal H}\alpha_{1_{\PS}}= \left( g + \frac{2 {\cal H} \dot {\cal D}}{3\mathcal{H}_0^2\Omega_{\rm m0}} \right)  \varphi_0
\end{equation}
is constant, as can be easily verified by taking the time derivative, and then using eq. \eqref{eqforD+} and the background Raychaudhuri equation $\dot{{\mathcal H}} - {\cal H}^2+ (3/2)\,{\cal H}^2\Omega_{\rm m}=0$. That constant can be obtained by evaluating eq.\,(\ref{eq:const}) at initial time. We have thus
\be \label{eq:constIN}
\varphi-{\cal H}\alpha_{1_{\PS}} = \frac 5 3 \varphi_{\rm in} \,, 
\ee
where $\varphi_{\rm in}=g_{\rm in} \varphi_0$. Then, our final expression for $\phi_{1_{\rm S}}$ is
\begin{equation} \label{phi1sfinal}
\phi_{1_{\rm S}}=\dfrac{2}{9} \dfrac{{\cal D}\nab^2\varphi_0}{\mathcal{H}_0^2\Omega_{\rm m0}}+\frac{5}{3}g_{\rm in}\varphi_0\,.
\end{equation}

Finally, by putting the above results together, we obtain the first-order result of the spatial metric
\be
 g_{ij_{\rm S}} = a^2  \left[ \left(1- \frac{10}{3} \varphi_{\rm in} \right) \delta_{ij} - \frac 4 3 \frac{{\cal D}}{\mathcal{H}_0^2\Omega_{\rm m0}} \varphi_{0,ij} \right] \,.
\ee

\paragraph{Four-velocity} The transformation of the four-velocity gives simply $u^\mu_{1_{\rm S}}= (1/a) (1,\fett{0})$, which is expected since the peculiar velocity (both temporal and spatial parts) vanishes by definition in the synchronous-comoving gauge.

\paragraph{Matter density} 
Finally, the transformation of the matter density perturbation in eq.\,\eqref{rho1} gives for the density contrast
\begin{equation} \label{delta1Ns}
\delta_{1_{\rm S}} =%D(\tau)\, C(\fett{q})=
  \frac{2}{3} \frac{{\cal D}\nab^2\varphi_0}{\mathcal{H}_0^2\Omega_{\rm m0}}\,,
\end{equation}
i.e., the first-order density contrast in the synchronous-comoving gauge is the solution of the Newtonian equation
\begin{equation}
\label{evdelta1}
\ddot{\delta}_{1} + {\cal H} \dot{\delta}_{1} - \frac{3}{2} \frac{{\cal H}_0^2 \Omega_{\rm m0}}{a} \delta_{1} =0\,. 
\end{equation} 
as it is well-known. The first-order relativistic correction in the Poisson gauge, which arises because of the non-vanishing velocity, is eliminated by the time transformation to the comoving (i.e., Lagrangian) frame.

All of our first-order results agree with those of refs.~\cite{BMPR2010,BHMW14} for $\Lambda$CDM.

\subsection{Second-order transformations}
\paragraph{Metric tensor: scalar perturbations}
From the transformation rules at second order \eqref{psi2}--\eqref{E2} we find
\begin{align}
0 &= \psi_{2_{\rm P}} + {\cal H} \alpha_{2_{\PS}} + \dot \alpha_{2_{\PS}} + \Pi_{{}_\PS} \,, \label{psi2s} \\ 
\phi_{2_{\rm S}} &= \phi_{2_{\rm P}} - {\cal H} \alpha_{2_{\PS}} - \frac{1}{6} \Upsilon^k_{\ k_{\PS}} -\frac{1}{3}\nab^2\beta_{2_{\PS}} \,, \label{phi2s}\\ 
0 &= - \alpha_{2_{\PS}} + \dot \beta_{2_{\PS}} + \nab^{-2} \Sigma^{\phantom{k},k}_{k_{\PS}} \,, \label{B2s} \\
E_{2_{\rm S}} &= \beta_{2_{\PS}} + \frac{3}{4} \nab^{-2} \nab^{-2} \Upsilon^{\phantom{ij},ij}_{ij_{\PS}} - \frac{1}{4} \nab^{-2} \Upsilon^{k}_{\ k_{\PS}} \,, \label{E2s} 
\end{align}
where the explicit expressions for the first-order squared terms $\Pi_{{}_\PS}$, $\Sigma_{i_{\PS}}$ and $\Upsilon_{ij_{\PS}}$ can be found in Appendix \ref{app:FOsquaredtermsPS}, whereas
 $\psi_{2_{\rm P}}$ and $\phi_{2_{\rm P}}$ are given in eqs.\,(\ref{eq:psi2P}) and~(\ref{eq:phi2P}). From eqs.\,(\ref{psi2s}) and~(\ref{B2s}), it is then straightforward to obtain the following expressions for the gauge generators
\begin{align}
 \alpha_{2_{\PS}} &= -\frac{1}{a} \int_{\eta_{\rm in}}^\eta \dd \tilde{\eta} \, a  \Big[ \psi_{2_{\rm P}}   +\Pi_{{}_\PS}  \Big] \,, \\
 \beta_{2_{\PS}} &=  \int_{\eta_{\rm in}}^\eta \dd \tilde \eta \left[ \alpha_{2_{\PS}}
      - \nab^{-2} \Sigma^k_{\ ,k_{\PS}} \right] \,. \label{beta2s}
\end{align}
To keep track of the computational steps, let us split $\alpha_{2_{\PS}}$ into a ``Newtonian''-like  and ``GR''-like part, i.e., $\alpha_{2_{\PS}} = \alpha_{2_{\PS}}^{\rm N}+ \alpha_{2_{\PS}}^{\rm GR}$. Here, with ``Newtonian''-like contributions, we mean that $\alpha_{2_\PS}^{\rm N}$ is identical with the Lagrangian velocity potential.\footnote{For the Newtonian non-perturbative approach to this gauge transformation see \cite{Villa:2014aja}.} Indeed
for the Newtonian part we have after some manipulations
\be
\alpha^{\rm N}_{2_{\PS}} =\frac{1}{a} \frac{4\Psi_0}{3\mathcal{H}_0^2\Omega_{\rm m0}}\int_{\eta_{\rm in}}^\eta \dd \tilde{\eta} \, \left({\cal D}^2 + {\cal F}\right) = \frac{8 \dot{\cal F}\Psi_0}{9\left(\mathcal{H}_0^2\Omega_{\rm m0}\right)^2} \,,
\ee
where the last expression is found by rewriting the differential equation~(\ref{eqF}) for the second-order growth function ${\cal F}$ in integral form:
\be
  \dot{\cal F} =\frac{3}{2} \frac{{\cal H}_0^2\Omega_{\rm m0}}{a}\int_{\eta_{\rm in}}^\eta \dd\tilde{\eta} \,({\cal D}^2 + {\cal F}) \,.
\ee
For the GR part of $\alpha_{2_{\PS}}$ we obtain after some straightforward calculations
\be \label{alpha2sGR}
 \alpha_{2_{\PS}}^{\rm GR} = \frac 1 a \int_{\eta_{\rm in}}^\eta \dd \tilde \eta
   \left[ \frac{10}{3} {\cal D} g_{\rm in} (a_{\rm nl} -1) \varphi_0^2 
  -8 \frac{{\cal H}^2 {\cal D}^2}{{\cal H}_0^2\Omega_{\rm m0}} \left( f^2 -f + \frac 3 2 \Omega_{\rm m}\right) \Theta_0 \right] \,.
\ee
To arrive at the terms proportional to the spatial kernel $\Theta_0$ we have used the relations~(\ref{omegam}) and~(\ref{Halpha}), and the fact that $\dot {\cal D} = f{\cal H}{\cal D}$.
The integral proportional to $\varphi_0^2$ is trivially solved by using~(\ref{magic4D}). To solve the other part of the integral~(\ref{alpha2sGR}), we note that
\be \label{DDdot}
  \frac{{\cal D}\dot{\cal D}}{a} =  \frac 1 a \int_{\eta_{\rm in}}^\eta \dd \tilde \eta\, {\cal H}^2{\cal D}^2\left(f^2-f+\frac 3 2 \Omega_{\rm m}\right) \,,
\ee
which can be easily proven by applying the ``reverted'' partial integration rule
\be
  \frac 1 a \left[ {\cal D} \dot {\cal D} \right] \Big|_{\eta_{\rm in}}^\eta =
     \frac 1 a \int_{\eta_{\rm in}}^\eta \dd \tilde \eta\, \left( \dot {\cal D}^2 + {\cal D} \ddot {\cal D} \right) \,,
\ee
discarding a decaying mode $\sim 1/a$, and by using the differential equation~(\ref{eqforD+}) to get an expression for $\ddot {\cal D}$. Putting the Newtonian and GR part of the temporal gauge generator together, we then obtain
\be
 \alpha_{2_{\PS}} = \frac{2\dot{{\cal D}}}{3\mathcal{H}_0^2\Omega_{\rm m0}}\left[ \frac{10}{3}g_{\rm in}\left(a_{\rm nl}-1\right) \varphi_0^2- 12g \Theta_0\right]
 + \frac{8 \dot{\cal F}\Psi_0}{9\left(\mathcal{H}_0^2\Omega_{\rm m0}\right)^2} \,.
\ee
Now we proceed with the calculation of the spatial gauge generator~(\ref{beta2s}), which we again split into a Newtonian and GR part, $\beta_{2_{\PS}}= \beta_{2_{\PS}}^{\rm N}+ \beta_{2_{\PS}}^{\rm GR}$. For the Newtonian part, the integration in time is trivial so we obtain immediately
\begin{align} \label{betaN}
  \beta^{\rm N}_{2_{\PS}} &= \frac{8}{9\left(\mathcal{H}_0^2\Omega_{\rm m0}\right)^2} \left[ {\cal F} \Psi_0- \frac{1}{4}{\cal D}^2 \left(\nab \varphi_0\right)^2 \right] \,,
\end{align}
where the last term originates from the Newtonian part of $\Sigma_{i_{\PS}}$, given in~(\ref{sigmaS}). For the GR part of $\beta_{2_{\PS}}$ we obtain
\begin{align} \label{beta2GRPS}
 \beta^{\rm GR}_{2_{\PS}} &= \frac{4}{3 \mathcal{H}_0^2\Omega_{\rm m0}} \int_{\eta_{\rm in}}^\eta \dd \tilde\eta \left[ \dot {\cal D} g_{\rm in} \left( \frac 5 3 a_{\rm nl} - \frac 5 2 \right) \varphi_0^2 - \dot {\cal D} g \varphi_0^2  - 6 \dot {\cal D} g \Theta_0 \right] \,.
\end{align}
The time integral of the first term is trivial, and to solve for the latter we
observe that
\be \label{intgDdot}
 \int_{\eta_{\rm in}}^\eta \dd \tilde \eta \dot {\cal D} g = \frac{\dot {\cal D}^2}{6 \mathcal{H}_0^2\Omega_{\rm m0}} + \frac 5 6 {\cal D} g_{\rm in} \,,
\ee
which can be obtained from the identity $\dot {\cal D}^2 \equiv 2 \int \dd \tilde \eta \dot {\cal D} \ddot {\cal D}$ and then using~(\ref{eqforD+}) and~(\ref{Halpha}).
We then obtain the GR part of the spatial gauge generator
\be
  \beta^{\rm GR}_{2_{\PS}}  =   \frac{2}{9 \mathcal{H}_0^2\Omega_{\rm m0}} \left[ 10 {\cal D} g_{\rm in} \left( a_{\rm nl} -2 \right) - \frac{\dot {\cal D}^2}{\mathcal{H}_0^2\Omega_{\rm m0}}  \right] \varphi_0^2
      - \frac{4}{3 \mathcal{H}_0^2\Omega_{\rm m0}} \left[ \frac{\dot {\cal D}^2}{\mathcal{H}_0^2\Omega_{\rm m0}} + 5 {\cal D} g_{\rm in} \right] \Theta_0 \,,
\ee
where we have set an integration constant to zero because we require that the Eulerian and Lagrangian frame coincide at initial time. 

We have thus solved for $\beta_{2_{\PS}} = \beta^{\rm N}_{2_{\PS}} + \beta^{\rm GR}_{2_{\PS}}$.
Equipped with the (longitudinal) gauge generators, we are now prepared to calculate the scalar perturbations in the synchronous-comoving gauge. We begin with the expression~(\ref{phi2s}) for the scalar $\phi_{2_{\rm S}}$. Again we split the quantitites into a Newtonian and GR part, $\phi_{2_{\rm S}}= \phi_{2_{\rm S}}^{\rm N}+ \phi_{2_{\rm S}}^{\rm GR}$. The only Newtonian part on the RHS of~(\ref{phi2s}) arises from $\beta_{2_{\PS}}^{\rm N}$ and some terms in
$\Upsilon_{ij_{\PS}}$, where the latter is given in eq.\,(\ref{upsilonS}). We thus have
\begin{align}
\phi^{\rm N}_{2_{\rm S}}&= -\frac{4}{27\left(\mathcal{H}_0^2\Omega_{\rm m0}\right)^2}\left[\left({\cal D}^2+{\cal F}\right) \varphi_{0,kl} \varphi_0^{,kl}-{\cal F}\left(\nab^2\varphi_0\right)^2\right] \,.
\intertext{For the GR part we obtain after some tedious but straightforward calculations}
 \phi_{2_{\rm S}}^{\rm GR} &= - \frac{50}{9} g_{\rm in}^2 a_{\rm nl} \varphi_0^2 + \frac{10}{9\mathcal{H}_0^2\Omega_{\rm m0}} {\cal D} g_{\rm in} \big( 1 - \frac 4 3 a_{\rm nl} \big) (\nab \varphi_0)^2
  +\frac{40 {\cal D} g_{\rm in}}{27 \mathcal{H}_0^2\Omega_{\rm m0}}  (1-a_{\rm nl}) \varphi_0 \nab^2 \varphi_0 \,.
\end{align}
To arrive at this result we have used the fact that
\be
  \frac{2 {\cal H} \dot {\cal F}}{\mathcal{H}_0^2\Omega_{\rm m0}} - \frac{\dot {\cal D}^2}{\mathcal{H}_0^2\Omega_{\rm m0}} - 5 {\cal D} g_{\rm in} + 3 \frac{{\cal D}^2+{\cal F}}{a} \equiv C_{\rm in}
\ee
is constant in time (and of course in space) which can be easily shown by taking the time derivative of the above and using the differential equations for the first-order and second-order growth functions, i.e., eqs.\,(\ref{eqforD+}) and~(\ref{eqF}). A careful
analysis reveals that $C_{\rm in}$ is vanishing.

Now let us proceed with the derivation of $E_{2_{\rm S}}= E_{2_{\rm S}}^{\rm N}+E_{2_{\rm S}}^{\rm GR}$.
From eq.\,(\ref{E2s}) we obtain for the Newtonian part directly
\begin{align}
 E_{2_{\rm S}}^{\rm N} &=  \frac{8}{9\left(\mathcal{H}_0^2\Omega_{\rm m0}\right)^2} {\cal F} \Psi_0
   + \frac{2 {\cal D}^2}{3 \left(\mathcal{H}_0^2\Omega_{\rm m0}\right)^2} \nab^{-2} \nab^{-2} D_{ij} \left( \varphi_{0,ik} \varphi_{0,j}^{,k} \right) \,.
\intertext{To solve for the GR part, we obtain firstly}
 E_{2_{\rm S}}^{\rm GR} &= \frac{2}{9 \mathcal{H}_0^2\Omega_{\rm m0}} \left[ 10 {\cal D} g_{\rm in} \left( a_{\rm nl} -2 \right) - \frac{\dot {\cal D}^2}{\mathcal{H}_0^2\Omega_{\rm m0}}  \right] \varphi_0^2
      - \frac{4}{3 \mathcal{H}_0^2\Omega_{\rm m0}} \left[ \frac{\dot {\cal D}^2}{\mathcal{H}_0^2\Omega_{\rm m0}} + 5 {\cal D} g_{\rm in} \right] \Theta_0 \nonumber \\
 &+  \nab^{-2} \nab^{-2}\left[ \frac{20}{3} \frac{g_{\rm in} {\cal D}}{\mathcal{H}_0^2\Omega_{\rm m0}}  + \frac 2 3 \frac{\dot {\cal D}^2}{(\mathcal{H}_0^2\Omega_{\rm m0})^2}  \right] \left( \varphi_{0} \varphi_{0}^{,ij} \right)_{,ij} \nonumber \\
&- \frac{20}{9}  \frac{g_{\rm in} {\cal D}}{\mathcal{H}_0^2\Omega_{\rm m0}} \nab^{-2} (\varphi_{0} \nab^2 \varphi_0)  - \frac 2 9   \frac{\dot {\cal D}^2}{(\mathcal{H}_0^2\Omega_{\rm m0})^2}  \nab^{-2} (\varphi_{0} \nab^2 \varphi_{0}) \,.  \label{E2Sfirst}
\end{align}
This expression can be drastically reduced by using the identity
\be
  \nab^{-2} \nab^{-2} D_{ij} \left( \varphi_{0} \varphi_{0}^{,ij} \right)
  =
   \frac{\varphi_0^2}{3} + 2 \Theta_0 \,,
\ee
and the relation \eqref{relationTheta0}.
We then obtain
\be
  E_{2_{\rm S}}^{\rm GR}  = \frac{20}{9 \mathcal{H}_0^2\Omega_{\rm m0}} {\cal D} g_{\rm in} \left[  \left( a_{\rm nl} -1 \right)  \varphi_0^2
     + 3 \Theta_0  \right] \,,
\ee
which concludes the derivation of $E_{2_{\rm S}}$.

\paragraph{Metric tensor: vector perturbations} 
From the second-order gauge transformations~(\ref{omega2}) and~(\ref{F2}) we obtain
\bea
 0 &=& \omega_{2i_{\rm P}} + \dot d_{2i_{\PS}}  \,, \\ 
F_{2i_{\rm S}} &=&  d_{2i_{\PS}} + \nab^{-2} \Upsilon_{ik_{\PS}}^{\phantom{ik},k} - \nab^{-2} \nab^{-2} \Upsilon^{\phantom{kl},kl}_{kl,i_{\PS}} \,,  \label{F2S} 
\eea
where the solutions for $\omega_{2i_{\rm P}}$ and $\Upsilon_{ij_{\PS}}$ are given in~(\ref{omega2ip}) and~(\ref{upsilonS}), respectively.
For the first equation we have used the fact that $\Sigma_{i_{\PS}} - \nab^{-2} \Sigma^{\phantom{k},k}_{k,i_{\PS}} =0$, where $\Sigma_{i_{\PS}}$ is given in~(\ref{sigmaS}).
The first equation gives then immediately the vector part of the gauge transformation
\begin{align}
  d_{2i_{\PS}}  &= \int_{\eta_{\rm in}}^\eta \dd \tilde \eta \frac{16}{3{\cal H}_0^2\Omega_{\rm m0}} \frac{{\cal D} \dot {\cal D}}{a} {\cal R}_i \nonumber \\
  &=   \frac{8}{9 {\cal H}_0^2\Omega_{\rm m0}} \left[ \frac{\dot {\cal D}^2}{{\cal H}_0^2\Omega_{\rm m0}} + 5 {\cal D} g_{\rm in} \right] {\cal R}_i \,,
\end{align}
where the transverse kernel ${\cal R}_i$ is given in~(\ref{transR}). We have used~(\ref{intgDdot}) to solve for the time integral, and, as above, an integration constant was set to zero by requiring that at initial time the
Eulerian and Lagrangian spatial coordinates coincide. Note that $ d_{2i_{\PS}}$ is purely relativistic.

Now we can easily derive from~(\ref{F2S}) the Newtonian and GR part of $F_{2i_{\rm S}}= F_{2i_{\rm S}}^{\rm N} + F_{2i_{\rm S}}^{\rm GR}$. They read
\begin{align}
  F_{2i_{\rm S}}^{\rm N} &=  \frac{8}{9} \frac{{\cal D}^2}{(\mathcal{H}_0^2\Omega_{\rm m0})^2}\left[ \nab^{-2}  (\varphi_{0,il} \varphi_{0}^{,kl})_{,k}
   - \nab^{-2} \nab^{-2} (\varphi_{0,m}^{,k} \varphi_{0}^{,lm})_{,kli} \right] \,,
   \intertext{and}
   F_{2i_{\rm S}}^{\rm GR} 
 &= - \frac{40}{9 {\cal H}_0^2\Omega_{\rm m0}} {\cal D} g_{\rm in} {\cal R}_i  \,,
\end{align}
where we have used the definition $\Theta_0 = -(1/2)\nab^{-2} \nab^{-2} D^{kl}(\varphi_{0,k} \varphi_{0,l})$, cf.~with~(\ref{Theta0}), and the relation \eqref{relationTheta0}.

\paragraph{Metric tensor: tensor perturbations}
We finally derive the tensor modes in the metric from the transformation rule~(\ref{chi2}). As usual, we split $\chi_{2 ij_{\rm S}} = \chi_{2 ij_{\rm S}}^{\rm N} + \chi_{2 ij_{\rm S}}^{\rm GR}$, and obtain for the respective parts
\begin{align}
% delta ij part:
 \frac 1 2 \chi_{2 ij_{\rm S}}^{\rm N} &= \frac{2{\cal D}^2}{9({\cal H}_0^2\Omega_{\rm m0})^2} \left[ - \varphi_{0,kl} \varphi_{0}^{,kl} + \nab^{-2} ( \varphi_{0}^{,kl} \varphi_{0,k}^{,m})_{,lm}  \right] \delta_{ij}  \nonumber \\
% ,ij part:
&+ \frac 2 9 \frac{{\cal D}^2}{({\cal H}_0^2\Omega_{\rm m0})^2} \left[ \nab^{-2} (\varphi_{0,kl} \varphi_{0}^{,kl})_{,ij} - 2 \nab^{-2} (\varphi_{0,il} \varphi_{0}^{,kl})_{,kj} -2 \nab^{-2} (\varphi_{0,jl} \varphi_{0}^{,kl})_{,ki} \right. \nonumber \\
   &\qquad+ \left. \nab^{-2} \nab^{-2} (\varphi_{0}^{,km} \varphi_{0,m}^{,l})_{,klij} 
  +2 \varphi_{0,ik} \varphi_{0,j}^{,k}  \right] \,, \\
  \frac 1 2 \chi_{2 ij_{\rm S}}^{\rm GR} &= \left( \frac{40}{9 {\cal H}_0^2\Omega_{\rm m0}} {\cal D} g_{\rm in} + \frac{8 \dot {\cal D}^2}{9({\cal H}_0^2\Omega_{\rm m0})^2} \right) \nab^{-2} {\cal S}_{ij}  + \frac 1 2 \pi_{2 ij_{\rm P}}   \,,
\end{align}
where $\pi_{2 ij_{\rm P}}$ is given in Eq.\,(\ref{GW}),
and we repeat the traceless and divergenceless tensor ${\cal S}_{ij}$ here for convenience
\be \label{SijRepeated}
  {\cal S}_{ij}(\fett{x}) = \nab^2 \Psi_0 \delta_{ij} +\Psi_{0,ij} +2 (\varphi_{0,ij}\nab^2 \varphi_0 - \varphi_{0,ik}\varphi_{0,j}^{,k} ) \,.
\ee
We shall comment on the tensors in section~\ref{summarySynch}, when we summarise our results in the synchronous-comoving gauge.

In concluding this section, let us emphasise that all the second-order synchronous-comoving gauge expressions obtained here are new. Only a few terms in these expressions were already known in the literature for the restricted case of an EdS universe (e.g., ref.~\cite{MMB}).

\paragraph{Four-velocity} As at first order, the scalar and transverse velocity vanishes in the Lagrangian frame by definition. 
\begin{align}
0&=v_{2_{\rm P}}-\dot \beta_{2_{\PS}}+\nab^{-2}\Omega^k_{\ ,k_{\PS}} \,, \label{v2s} \\ 
0&=w^i_{2_{\rm P}}-\dot d^i_{2_{\PS}}+\Omega^{i}_{{}_\PS}-\nab^{-2}\Omega^k_{\ ,ki_{\PS}}\,. \label{w2s}  
\end{align}
In both expressions, everything is already determined. We have used however these expressions to perform a consistency check of our calculations, e.g., from eq.\,(\ref{v2s}) we can rederive $v_{2_{\rm P}}$ (or, equivalently, by using the solutions of the RHS in~(\ref{v2s}) and~(\ref{w2s}), we can verify the validity of the comoving gauge conditions $v = w^i=0$ on the respective LHS's, as it should for an irrotational and pressure-less fluid).
For the perturbation of the zeroth component of the four velocity we obtain from eq.\,(\ref{v0any}) that
\be
 v^0_{2_{\rm S}} =0 \,.
\ee

\paragraph{Matter density} 
Finally, from the formula~(\ref{rho2}), we find the second-order density contrast $\delta_{2_{\rm S}}\equiv \delta_{2_{\rm S}}^{\rm N} +  \delta_{2_{\rm S}}^{\rm GR}$,
\begin{align}
 \delta_{2_{\rm S}}^{\rm N} &= \frac{4}{9 ({\cal H}_0^2 \Omega_{\rm m0})^2} \Bigg[ ({\cal D}^2+{\cal F}) (\nab^2\varphi_{0})^2 + ({\cal D}^2 - {\cal F}) \varphi_{0,lm} \varphi_0^{,lm} \Bigg] \,, \\
 \delta_{2_{\rm S}}^{\rm GR} &=    
    \frac{40 {\cal D} g_{\rm in}}{9{\cal H}_0^2 \Omega_{\rm m0}} \left[ \left( \frac 3 4 - a_{\rm nl} \right) (\nab \varphi_0)^2
    +  \left( 2 - a_{\rm nl} \right)  \varphi_0 \nab^2 \varphi_0 \right]  \,.
\end{align}
The Newtonian part has been reported in the GR literature in e.g.~\cite{Rampf:2013ewa}, which however relies on a different solution technique namely the gradient expansion.
The GR part, however, is fairly known see e.g., \cite{BMPR2010,Bruni:2014xma,Rampf:2014mga}.

%%%%%%%%%%%%%%%%%%%%%%%%%%%%%%%%%%%%%%%%%%%%%%%%%%%%%%%%%%%%%%%%%%%%%%%%%%%%%%%%%%%%%%%%%%%%%%%%%%%%%%%%%%%%%%%
\section{From the synchronous-comoving gauge to the total matter gauge}\label{synchro2x}
%%%%%%%%%%%%%%%%%%%%%%%%%%%%%%%%%%%%%%%%%%%%%%%%%%%%%%%%%%%%%%%%%%%%%%%%%%%%%%%%%%%%%%%%%%%%%%%%%%%%%%%%%%%%%

In this section we adopt the general formulas of section \ref{gaugetformulas} to the transformation from the synchronous-comoving gauge to the total matter gauge \cite{malik&wands} (sometimes also known as the velocity orthogonal gauge \cite{kodama}, or Eulerian gauge \cite{Rampf:2014mga}).

 We define the total matter gauge perturbatively with (see section~\ref{sec:lageul})
\begin{align}
  E_{(r)_{{\rm T}}} &= F_{(r)_{{\rm T}}} ^i = 0   \,, \qquad (\text{spatial~gauge~condition})  \label{spaceT} \\
  {\cal S}^i\,{T^0}_{i(r)_{{\rm T}}} &= 0\,, \qquad \hspace{1.35cm} (\text{temporal~gauge~condition})  \label{timeT}
\end{align}
where ${\cal S}^i = \nab^{-2} \partial^i$.
Relation~(\ref{timeT}) implies at first order (cf.\ eq.\,(\ref{T0i1}))
\be
 v_{1_{\rm T}}+B_{1_{\rm T}}=0\,, \label{currenttempTOM1}
\ee
and at second order (cf.\ eq.\,(\ref{T0i2}))
\be
 v_{2_{\rm T}} + B_{2_{\rm T}} -2 \nab^{-2} \partial^l \left( \psi_{1_{\rm T}} B_{1_{\rm T},l} + 2 \phi_{1_{\rm T}} v_{1_{\rm T},l}\right) = 0 \,. \label{currenttempTOM2}
\ee
As we shall show in the following, these conditions imply for an irrotational and pressure-less fluid that the time gauge generator of the gauge transformation from the synchronous-comoving gauge is vanishing at first and second order.

\subsection{First-order transformations} \label{FO_TOM}
\paragraph{Metric tensor} 
From transformation rules derived in section~\ref{generalFO}, we have for the
considered gauge transformation
\begin{align}
\psi_{1_{\rm T}} &= {\cal H}\alpha_{1_{\ST}}+\dot{\alpha}_{1_{\ST}} \,,  \label{psi1x}\\ 
\phi_{1_{\rm T}} &= \phi_{1_{\rm S}} -\frac{1}{3}\nab^2\beta_{1_{\ST}} \,, \label{phi1x}\\ 
B_{1_{\rm T}} &= \dot \beta_{1_{\ST}} \,, \label{B1x} \\
0 &= E_{1_{\rm S}} + \beta_{1_{\ST}} \label{E1x} \,.
\end{align}
Using the first-order result~(\ref{ourE1s}) for $E_{1_{\rm S}}$, 
equation \eqref{E1x} gives immediately for the spatial gauge generator 
\be \label{beta1x}
\beta_{1_{\ST}}= \frac{2}{3} \frac{{\cal D}\varphi_0}{\mathcal{H}_0^2\Omega_{\rm m0}} \,.
\ee
Note that, apart from the minus sign, this spatial generator is the same as from the Poisson gauge to the synchronous-comoving. The minus sign arises because in this section we perform the transformation from Lagrangian to Eulerian frame, whereas in section~\ref{Poisson} it was the other way around.

To obtain the temporal gauge generator $\alpha_{1_{\rm ST}}$, we sum up eqs.\,(\ref{B1}) and~(\ref{v1}) which gives the general expression $\tilde v_1 + \tilde B_1 = v_1 + B_1 - \alpha_1$. For the considered gauge transformation, where $v_{1_{\rm S}}= B_{1_{\rm S}}=0$ and because of the temporal gauge condition~(\ref{currenttempTOM1}),
this implies
\be
  \alpha_{1_{\rm ST}} = 0 \,.
\ee
Since we assume vanishing first-order vector perturbations, we have $d^i_{1_{\rm ST}}=0$.
Having determined the gauge generators, the remaining first-order scalar perturbations
are easily derived,
\begin{align}
 \psi_{1_{\rm T}} &= 0 \,, \\
 \phi_{1_{\rm T}} &= \frac{5}{3}\varphi_{\rm in} \,, \\ 
  B_{1_{\rm T}} &= \frac{2}{3} \frac{\dot{\cal D}\varphi_0}{\mathcal{H}_0^2\Omega_{\rm m0}} \,. 
\end{align}

\paragraph{Four-velocity} 
The first-order spatial peculiar velocity in the total matter gauge is given by
\begin{equation}\label{v1x}
  {v}_{1_{\rm T}}=-\dot \beta_{1_{\ST}} = - \frac{2}{3} \frac{\dot {\cal D}\varphi_0}{\mathcal{H}_0^2\Omega_{\rm m0}} \,,
\end{equation}
which follows directly from~(\ref{v1}). Note that the spatial peculiar velocity \eqref{v1x} is identical to the one in the Poisson gauge (cf.~(\ref{v1P})).
For the perturbation of the zeroth component of the four velocity, since $\psi_{1_{\rm T}}=0$ we obtain from eq.\,(\ref{v0any}) that
\be
 v^0_{1_{\rm T}} =0 \,.
\ee

\paragraph{Matter density}  The matter density is identical to the one in the synchronous-comoving gauge because the time coordinate is identical and the matter density is unaffected by a purely spatial transformation at first order.
We thus have directly from \eqref{rho1}: 
\begin{equation}\label{rho1x}
\delta_{1_{\rm T}}= \delta_{1_{\rm S}} =  \frac{2}{3} \frac{{\cal D}\nab^2\varphi_0}{\mathcal{H}_0^2\Omega_{\rm m0}}\,.
\end{equation}

\subsection{Second-order transformations}

\paragraph{Metric tensor: scalar perturbations}

From the transformation rules at second order \eqref{psi2}--\eqref{E2} we have
\begin{align}
 \psi_{2_{\rm T}} &=  {\cal H} \alpha_{2_{\ST}} + \dot \alpha_{2_{\ST}} + \Pi_{{}_\ST} \,,  \label{psi2T} \\
 \phi_{2_{\rm T}} &= \phi_{2_{\rm S}} - \frac{1}{6} \Upsilon^k_{\ k_{\ST}} -\frac{1}{3}\nab^2\beta_{2_{\ST}}\,,  \label{phi2x}\\ 
 B_{2_{\rm T}} &=  \dot \beta_{2_{\ST}} + \nab^{-2} \Sigma^{\phantom{k},k}_{k_{\ST}} \,, \label{B2x} \\
0 &= E_{2_{\rm S}} + \beta_{2_{\ST}} + \frac{3}{4} \nab^{-2} \nab^{-2} \Upsilon^{\phantom{ij},ij}_{ij_{\ST}} - \frac{1}{4} \nab^{-2} \Upsilon^{k}_{\ k_{\ST}} \label{E2x}  \,,
\end{align}
where the first-order squared terms $\Pi_{\rm ST}$, $\Upsilon_{ij_{\ST}}$, and $\Sigma_{i_{\ST}}$ are given in appendix~\ref{app:FOsquaredtermsST}.
In writing down these equations we have only used the spatial gauge condition for the total matter gauge, see eq.\,(\ref{spaceT}). The temporal gauge condition~(\ref{currenttempTOM2}) of the total matter gauge becomes at second order
\be \label{defx2metric}
 v_{2_{\rm T}} + B_{2_{\rm T}} = 2 \nab^{-2} \partial^l \left( \psi_{1_{\rm T}} B_{1_{\rm T},l} + 2 \phi_{1_{\rm T}} v_{1_{\rm T},l}\right) = - \frac{20 \dot {\cal D} g_{\rm in}}{9 \stuff}  \varphi_0^2 \,,
\ee
where we have used the first-order results from section~\ref{FO_TOM}. 
Now, to obtain the temporal gauge generator $\alpha_{2_{\rm ST}}$, we sum up eqs.\,(\ref{B2}) and~(\ref{v2}) which yields the general expression $\tilde v_2 + \tilde B_2 = v_2 + B_2 + \nab^{-2} {\Omega^k}_{,k}- \alpha_2 + \nab^{-2} \Sigma^{\phantom{k},k}_{k}$. For the considered gauge transformation, where $v_{2_{\rm S}}= B_{2_{\rm S}}=\Omega_{\rm ST}^k =0$ and $\nab^{-2} \Sigma^{\phantom{k},k}_{k_{\rm ST}}= - \frac{20 \dot {\cal D} g_{\rm in}}{9 \stuff}  \varphi_0^2$  (see appendix~\ref{app:FOsquaredtermsST}), we obtain for the time gauge generator at second order
\be
  \alpha_{2_{\ST}} = 0 \,.
\ee
Also the derivation of the longitudinal part of the spatial gauge generator is straightforward.
From eq.\,(\ref{E2x}) we get
\begin{align}
 \beta_{2_{\ST}}^{\rm N} &= - \frac 8 9 \frac{{\cal F}}{({\cal H}_0^2 \Omega_{\rm m0})^2} \Psi_0
   + \frac 2 9 \frac{{\cal D}^2}{({\cal H}_0^2 \Omega_{\rm m0})^2} \varphi_{0,l} \varphi_{0}^{,l} \,, \\
 \beta_{2_{\ST}}^{\rm GR} &=  - \frac{20}{9 \mathcal{H}_0^2\Omega_{\rm m0}} {\cal D} g_{\rm in} \left( a_{\rm nl} -2 \right)  \varphi_0^2
      + \frac{20}{3 \mathcal{H}_0^2\Omega_{\rm m0}} {\cal D} g_{\rm in} \Theta_0 \,,
\end{align}
where $\beta_{2_{\ST}} \equiv \beta_{2_{\ST}}^{\rm N} + \beta_{2_{\ST}}^{\rm GR}$.
The Newtonian part is, of course apart from a global minus sign, identical with the Newtonian part of $\beta_{2_{\PS}}$, see eq.\,(\ref{betaN}).
The GR part is identical to $\beta_{2_\PS}^{\rm GR}$ (see eq.\,(\ref{beta2GRPS})), apart from the terms $\sim \dot {\cal D}^2$, which arise from the different time coordinate used.

The calculation of the other metric perturbations is now trivial. We obtain
\begin{align}
  \psi_{2_{\rm T}} &= - v_{1_{\rm T},l} v_{1_{\rm T}}^{,l}  = - \frac 4 9 \frac{\dot {\cal D}^2}{(\stuff)^2}  \varphi_{0,l} \varphi_{0}^{,l} \,, \\
  \phi_{2_{\rm T}} &\equiv \phi_{2_{\rm T}}^{\rm GR} = - \frac{50}{9} g_{\rm in}^2 a_{\rm nl} \varphi_0^2
   + \frac{10}{9} \frac{{\cal D} g_{\rm in}}{{\cal H}_0^2 \Omega_{\rm m0}} \varphi_{0,l} \varphi_{0}^{,l} - \frac{20}{9} \frac{{\cal D} g_{\rm in}}{{\cal H}_0^2 \Omega_{\rm m0}} \Psi_0  \,, \\
  B_{2_{\rm T}} &\equiv B_{2_{\rm T}}^{\rm N} + B_{2_{\rm T}}^{\rm GR} \,,
\intertext{where}
 B_{2_{\rm T}}^{\rm N} &=- \frac 8 9 \frac{\dot {\cal F}}{({\cal H}_0^2 \Omega_{\rm m0})^2} \Psi_0
   + \frac 4 9 \frac{{\cal D} \dot {\cal D}}{({\cal H}_0^2 \Omega_{\rm m0})^2} \varphi_{0,l} \varphi_{0}^{,l}  \,, \label{Bnewtonian}\\
 B_{2_{\rm T}}^{\rm GR}  &=  - \frac{20}{9 \mathcal{H}_0^2\Omega_{\rm m0}} \dot {\cal D} g_{\rm in} \left( a_{\rm nl} -1 \right)  \varphi_0^2
      + \frac{20}{3 \mathcal{H}_0^2\Omega_{\rm m0}} \dot {\cal D} g_{\rm in} \Theta_0 \,.
\end{align}
Note that eq.\,(\ref{Bnewtonian}) is the second-order peculiar velocity potential in NEPT (see eq.\,(\ref{eqv2p})).

\paragraph{Metric tensor: vector perturbations}
From the second-order gauge transformations~(\ref{omega2}) and~(\ref{F2}) we obtain
\bea
 \omega_{2i_{\rm T}} &=&  \dot d_{2i_{\ST}} +\Sigma_{i_{\ST}} - \nab^{-2} \Sigma^{\phantom{k},k}_{k,i_{\ST}}  \,, \label{omega2x} \\
0 &=& F_{2i_{\rm S}} + d_{2i_{\ST}} + \nab^{-2} \Upsilon^{\phantom{ki},k}_{ki_{\ST}} - \nab^{-2} \nab^{-2} \Upsilon^{\phantom{kl},kl}_{kl, i_{\ST} } \,, \label{F2x} 
\eea
where $\Upsilon_{ij_{\ST}}$ and $\Sigma_{i_{\ST}}$ are given in eqs.~(\ref{Sx}) and~(\ref{Ux}).
Again, the calculations are simple so we leave them as an exercise. We obtain
\begin{align}
d_{2i_{\ST}} &\equiv F_{2i_{\rm S}}^{\rm GR} =  - \frac{40}{9 {\cal H}_0^2\Omega_{\rm m0}} {\cal D} g_{\rm in} \,   {\cal R}_i \,, \\
w_{2i_{{\rm T}}} &\equiv \dot d_{2i_{\ST}} =  - \frac{40}{9 {\cal H}_0^2\Omega_{\rm m0}} \dot {\cal D} g_{\rm in}  \, {\cal R}_i \,,
\end{align}
where ${\cal R}_i$ is given in eq.\,(\ref{transR}).

\paragraph{Metric tensor: tensor perturbations}
We obtain this simple result
\begin{align}
 \chi_{2ij_{\rm T} } &= %\frac 8 9 \frac{\dot D^2}{({\cal H}_0^2 \Omega_{m0})^2} \nab^{-2} \left[ \delta_{ij} \nab^2 \Psi_0 +  \Psi_{0,ij} + 2 \left( \varphi_{0,ij} \varphi_{0,ll} - \varphi_{0,il} \varphi_{0,jl} \right)    \right] + \chi^{{\rm T}}_{2_P ij} \nonumber \\  &\equiv 
\frac 8 9 \frac{\dot {\cal D}^2}{({\cal H}_0^2 \Omega_{\rm m0})^2} \nab^{-2}  {\cal S}_{ij} + \pi_{2 ij_{\rm P}} \,,
\end{align}
where ${\cal S}_{ij}$ is given in eq.\,(\ref{SijRepeated}), and $\pi_{2ij_{\rm P}}$, which contain propagating tensor modes (gravitational waves) in the Poisson gauge, can be found in~(\ref{GW}). Observe that the ``Newtonian-like'' tensors (growing with the same amplitude as second-order Newtonian perturbations), and partly the post-Newtonian tensor, that we have in the synchronous-comoving gauge, disappeared.

Let us emphasise again that all these $\Lambda$CDM results are new.

\paragraph{Four-velocity} The transformation rules~(\ref{v2})--(\ref{w2}) deliver simple expressions,
for the scalar and vector part of the spatial peculiar velocity we find firstly 
\begin{align}
 {v}_{2_{\rm T}}   &=-\dot \beta_{2_{\ST}} \,, \label{v2x} \\ 
 {w}_{2i_{\rm T}} &=-\dot d_{2i_{\ST}} \,, \label{w2x} 
\end{align}
from which we get
\begin{align}
 {v}_{2_{\rm T}}^{\rm N} &=  \frac 8 9 \frac{\dot {\cal F}}{({\cal H}_0^2 \Omega_{\rm m0})^2} \Psi_0
   - \frac 4 9 \frac{\dot {\cal D} {\cal D}}{({\cal H}_0^2 \Omega_{\rm m0})^2} \varphi_{0,k} \varphi_{0}^{,k} \,, \\
 {v}_{2_{\rm T}}^{\rm GR} &=  \frac{20}{9 \mathcal{H}_0^2\Omega_{\rm m0}} \dot {\cal D} g_{\rm in} \left( a_{\rm nl} -2 \right)  \varphi_0^2
      - \frac{20}{3 \mathcal{H}_0^2\Omega_{\rm m0}} \dot {\cal D} g_{\rm in} \Theta_0 \,, \\
 {w}_{2i_{\rm T}} &=   \frac{40}{9 {\cal H}_0^2\Omega_{\rm m0}} \dot {\cal D} g_{\rm in} {\cal R}_i \,,
\end{align}
where we split as usual ${v}_{2_{\rm T}} = {v}_{2_{\rm T}}^{\rm N}+ {v}_{2_{\rm T}}^{\rm GR}$. 
For the perturbation of the zeroth component of the four velocity we obtain from eq.\,(\ref{v0any}) that
\be
 v^0_{2_{\rm T}} =0 \,.
\ee

\paragraph{Matter density}
Finally, according to~(\ref{rho2}) the matter density perturbation transforms as
\begin{equation}\label{rho2x}
 {\delta}_{2_{\rm T}}= \delta_{2_{\rm S}} + 2\delta_{1_{\rm S},k}\xi_{1_{\ST}}^k\,.
\end{equation}
Since the last term on the RHS is purely Newtonian, 
the GR corrections are identical to the one in the synchronous-comoving gauge.\footnote{We note that it is in principle possible to define an Eulerian gauge where the
last term on the RHS of~(\ref{rho2x}) will not only contain a Newtonian part but also 
a relativistic part \cite{FidlerEtAll}.}
We get then for the density contrast 
\begin{align}
 \delta_{2_{\rm T}} &= \frac{4}{9 ({\cal H}_0^2 \Omega_{\rm m0})^2} \Bigg[ ({\cal D}^2+{\cal F}) (\nab^2\varphi_{0})^2 + 2{\cal D}^2 \varphi_{0,k} \nab^2 \varphi_{0}^{,k} + ({\cal D}^2 - {\cal F}) \varphi_{0,lm} \varphi_0^{,lm} \Bigg]  \nonumber \\ 
  &\qquad+ \frac{40 {\cal D} g_{\rm in}}{9{\cal H}_0^2 \Omega_{\rm m0}} \left[ \left( \frac 3 4 - a_{\rm nl} \right) (\nab \varphi_0)^2
    +  \left( 2 - a_{\rm nl} \right)  \varphi_0 \nab^2 \varphi_0 \right]  \,,
\end{align}
which agrees with the findings in \cite{Rampf:2014mga}.

%%%%%%%%%%%%%%%%%%%%%%%%%%%%%%%%%%%%%%%%%%%%%%%%%%%%%%%%%%%%%%%%%%%%%%%%%%%%%%%%%%%%%%%%%%%%%%%%%%%%%%%%%%%%%%%%%%%%
\section{Summary of the results}\label{summary}

In this section we summarise our results in the three gauges, namely 
the Poisson gauge~\ref{summaryPoisson}, the synchronous-comoving gauge~\ref{summarySynch} and the total matter gauge~\ref{summaryTom}. Here we only show the results for the 
metric and for the density contrast.
For the summary of the gauge generators see section~\ref{summaryGauge}, and 
 for more results we refer to the sections above. For an overview of the used notation, see tab.~\ref{tab:notation} in appendix~\ref{app:notation}.
We recall that for the EdS limit ($\Omega_{\rm m} = 1$) we have
\begin{align}
  \begin{split}
    a &= \eta^2 \,, \quad \qquad {\cal H} = \frac{2}{\eta} \,, \qquad\quad \stuff = 4 \,, \\
    {\cal D} &= \eta^2  \,, \quad \hspace{0.8cm}  {\cal F} = \frac 3 7 \eta^4   \,, \qquad \hspace{1.01cm}  g = g_{\rm in} =f=1  \,.
  \end{split}  \qquad \qquad \quad \text{(EdS universe)}
\end{align}

\subsection{Poisson gauge}\label{summaryPoisson}

We obtain for the metric components in the Poisson gauge
\begin{align}
  g_{00_{\rm P}} &= -a^2 \Bigg[ 1+ 2 g \varphi_0  +
\left( 3 g^2 + \frac 5 3 g g_{\rm in} (1- 2a_{\rm nl})  + \frac{2 \dot {\cal D}^2}{3a \mathcal{H}_0^2\Omega_{\rm m0}}  \right)\varphi_0^2 \nonumber \\ 
&\quad\hspace{0.5cm}+ 6 \bigg(4 g^2 - \frac{10}{3} g g_{\rm in} + \frac 4 3 \frac{\dot {\cal D}^2}{a \mathcal{H}_0^2\Omega_{\rm m0}} \bigg) \Theta_0 %\nonumber \\&\quad\hspace{2cm}
+ \frac{2{\cal D}^2}{3a\mathcal{H}_0^2\Omega_{\rm m0}}  \varphi_{0,l} \varphi_0^{,l}-\frac{4\left({\cal D}^2+{\cal F}\right)}{3a\mathcal{H}_0^2\Omega_{\rm m0}}\Psi_0   \Bigg] \,, \\
 g_{0i_{\rm P}}  &=  
   - \frac{8 a {\cal D} \dot {\cal D} }{3{\cal H}_0^2\Omega_{\rm m0}} {\cal R}_i \,,  \\
 g_{ij_{\rm P}}  &= a^2 \Bigg[ \delta_{ij} \bigg\{  1 - 2g \varphi_0
   + \left(  g^2 - \frac 5 3 g g_{\rm in} (1- 2a_{\rm nl})  - \frac{2 \dot {\cal D}^2}{3a \mathcal{H}_0^2\Omega_{\rm m0}}  \right)\varphi_0^2 %\nonumber \\&
- 6\left( 2 g^2 -\frac{10}{3}g g_{\rm in}  \right) \Theta_{0} \nonumber \\
 &\quad\hspace{2cm}-\frac{2{\cal D}^2}{3a\mathcal{H}_0^2\Omega_{\rm m0}} \varphi_{0,l} \varphi_0^{,l}+\frac{4\left({\cal D}^2+{\cal F}\right)}{3a\mathcal{H}_0^2\Omega_{\rm m0}}\Psi_0 
\bigg\} + \frac 1 2 \pi_{2 ij_{\rm P}} \Bigg] \,,  
\end{align}
where $\pi_{2 ij_{\rm P}}$ is given in eq.\,(\ref{GW}).
The density contrast is 
\begin{align}
  \delta_{\rm P} &= \frac{2}{3 {\cal H}_0^2 \Omega_{\rm m0}} \left[ {\cal D} \nab^2 \varphi_0 - 3 {\cal H} \dot {\cal D} \varphi_0 \right] \nonumber \\
&\qquad+ \frac{{\cal H}^2 {\cal D}^2}{a {\cal H}_0^2 \Omega_{\rm m0}} \left[ f^2 -4f \right] \varphi_0^2 + \frac{5 {\cal H}}{3{\cal H}_0^2 \Omega_{\rm m0}} (1+2a_{\rm nl}) \dot {\cal D} g_{\rm in} \varphi_0^2 -\frac{12 {\cal H} \dot {\cal D} {\cal D}}{a {\cal H}_0^2 \Omega_{\rm m0}} \Theta_0  \nonumber\\
&\qquad+ \frac{{\cal D}}{{\cal H}_0^2 \Omega_{\rm m0}} \left[ g - \frac{20}{9} g_{\rm in} a_{\rm nl}  \right] (\nab \varphi_0)^2 +\frac{4 {\cal H} \dot {\cal F}}{3({\cal H}_0^2 \Omega_{\rm m0})^2} \Psi_0\nonumber \\
&\qquad+  \frac{1}{3{\cal H}_0^2 \Omega_{\rm m0}}\left[ 6 {\cal D} g +\frac{10}{3} {\cal D} g_{\rm in} (1- 2a_{\rm nl}) + \frac 4 3 \frac{\dot {\cal D}^2}{{\cal H}_0^2 \Omega_{\rm m0}}  \right]  \varphi_0 \nab^2 \varphi_0 \nonumber \\
 &\qquad+ \frac{2}{9 ({\cal H}_0^2 \Omega_{\rm m0})^2} \Bigg[ ({\cal D}^2+{\cal F}) (\nab^2\varphi_{0})^2 + 2 {\cal D}^2 \varphi_{0}^{,l} \nab^2 \varphi_{0,l} + ({\cal D}^2 - {\cal F}) \varphi_{0,lm} \varphi_0^{,lm} \Bigg] \,.
\end{align}

\subsection{Synchronous-comoving gauge}\label{summarySynch}

We obtain for the metric components in the synchronous-comoving gauge
\begin{align}
  g_{00_{\rm S}} &= -a^2 \,, \\
  g_{0i_{\rm S}} &= 0 \,, \\
  g_{ij_{\rm S}} &=  a^2 \Bigg\{ \left( 1- \frac{10}{3} \varphi_{\rm in} + \frac{50}{9}  a_{\rm nl} \varphi_{\rm in}^2
   + \frac{10}{9{\cal H}_0^2\Omega_{\rm m0}} {\cal D} g_{\rm in} \varphi_{0,l} \varphi_{0}^{,l} \right) \delta_{ij}  - \frac 4 3 \frac{{\cal D}}{\mathcal{H}_0^2\Omega_{\rm m0}}  \varphi_{0,ij} 
  \Bigg. \nonumber \\
 &+ \frac{40}{9{\cal H}_0^2\Omega_{\rm m0}} {\cal D} g_{\rm in} \left[ \left(a_{\rm nl} - \frac 3 2 \right) \varphi_{0,i} \varphi_{0,j} +(a_{\rm nl} -1) \varphi_{0} \varphi_{0,ij} \right]  + \frac 1 2 \pi_{2ij_{\rm S}}   \nonumber \\ 
 & + \frac{4}{9({\cal H}_0^2\Omega_{\rm m0})^2} \left[ {\cal D}^2 \varphi_{0,ik} \varphi_{0,j}^{,k} -2 {\cal F} \nab^2 \Psi_0 \delta_{ij}
    -4{\cal F} (\varphi_{0,ij} \nab^2 \varphi_{0} - \varphi_{0,il} \varphi_{0,j}^{,l})
  \right] \Bigg. \Bigg.\Bigg\}
\,,
\end{align}
where the tensor
\begin{align} \label{solGWsynch6.50}
 \pi_{2ij_{\rm S}} =  \frac{16{\cal F}}{9({\cal H}_0^2\Omega_{\rm m0})^2} {\cal S}_{ij}
  + \left( \frac{40}{9{\cal H}_0^2\Omega_{\rm m0}} {\cal D} g_{\rm in} + \frac{8 \dot {\cal D}^2}{9({\cal H}_0^2\Omega_{\rm m0})^2}  \right) \nab^{-2} {\cal S}_{ij}   +  \pi_{2ij_{\rm P}} 
\end{align}
is the solution of the gravitational wave equation
\be \label{GWsynch6.51}
 \ddot \pi_{2ij_{\rm S}} + 2 {\cal H} \dot \pi_{2ij_{\rm S}} - \nab^2 \pi_{2ij_{\rm S}} = -\frac{16{\cal F}}{9({\cal H}_0^2\Omega_{\rm m0})^2} \nab^2 {\cal S}_{ij} \,.
\ee  
This result, valid for $\Lambda$CDM, is new and coincides with ref.~\cite{MMB} 
in the EdS limit.
Comparing the gravitational wave equation~(\ref{GWsynch6.51}) with the one obtained in the Poisson gauge (Eq.\,(\ref{waveeq})), it is evident that the nature of the wave equation has not changed but only its source term. The solution of the wave equation includes $ \pi_{2ij_{\rm P}}$, given in eq.\,(\ref{GW}), however,
as a consequence of the different source term, solution~(\ref{solGWsynch6.50}) contains also additional tensor perturbations that grow with ${\cal F}$, ${\cal D}$ and $\dot {\cal D}^2$. The physical interpretation of the various tensor perturbations is highly non-trivial but has been attempted in refs.~\cite{MMB,Rampf:2014mga}.

%As a consequence of the different source term, however,  the solution~(\ref{solGWsynch6.50}) contains also tensor perturbations of the Newtonian and post-Newtonian kind. For the physical interpretation of the various tensor perturbations, we refer to the discussions of refs.~\cite{MMB,Rampf:2014mga}.

For the density contrast we obtain 
\begin{align}
  \delta_{\rm S} &= \frac{2}{3 {\cal H}_0^2 \Omega_{\rm m0}}  {\cal D} \nab^2 \varphi_0
    + \frac{20 {\cal D} g_{\rm in}}{9{\cal H}_0^2 \Omega_{\rm m0}} \left[ \left( \frac 3 4 - a_{\rm nl} \right) (\nab \varphi_0)^2
    +  \left( 2 - a_{\rm nl} \right)  \varphi_0 \nab^2 \varphi_0 \right] \nonumber \\
  &\qquad+\frac{2}{9 ({\cal H}_0^2 \Omega_{\rm m0})^2} \Bigg[ ({\cal D}^2+{\cal F}) (\nab^2\varphi_{0})^2 + ({\cal D}^2 - {\cal F}) \varphi_{0,lm} \varphi_0^{,lm} \Bigg] \,.
\end{align}

%%%%%%%%%%%%%%%%%%%%%%%%%%%%%%
\subsection{Total matter gauge}\label{summaryTom}

We obtain for the metric components in the total matter gauge
\begin{align}
  g_{00_{\rm T}} &= -a^2 \left[ 1 - \frac{4}{9} \frac{\dot {\cal D}^2}{({\cal H}_0^2 \Omega_{\rm m0})^2} (\nab\varphi_{0})^2   \right]  \,, \\
 g_{0i_{\rm T}} &= a^2 \Bigg[  \frac{2}{3} \frac{\dot{\cal D}\varphi_{0,i}}{\mathcal{H}_0^2\Omega_{\rm m0}} 
  - \frac{20}{9 \mathcal{H}_0^2\Omega_{\rm m0}} \dot {\cal D} g_{\rm in} \left( a_{\rm nl} -1 \right)  \varphi_0 \varphi_{0,i}
      + \frac{10}{3 \mathcal{H}_0^2\Omega_{\rm m0}} \dot {\cal D} g_{\rm in} \Theta_{0,i} \nonumber \\ &\qquad 
 - \frac{20}{9 {\cal H}_0^2\Omega_{\rm m0}} \dot {\cal D} g_{\rm in}  \, {\cal R}_i  %\nonumber \\ &\qquad 
 - \frac 4 9 \frac{\dot {\cal F}}{({\cal H}_0^2 \Omega_{\rm m0})^2} \Psi_{0,i}
   + \frac 4 9 \frac{{\cal D} \dot {\cal D}}{({\cal H}_0^2 \Omega_{\rm m0})^2} \varphi_{0,li} \varphi_{0}^{,l}  
 \Bigg]  \,, \\
  g_{ij_{\rm T}} &= a^2 \Bigg[ \delta_{ij} \left( 1 - \frac{10}{3}\varphi_{\rm in}
   + \frac{50}{9} g_{\rm in}^2 a_{\rm nl} \varphi_0^2
   - \frac{10}{9} \frac{{\cal D} g_{\rm in}}{{\cal H}_0^2 \Omega_{\rm m0}} \varphi_{0,l} \varphi_{0}^{,l} + \frac{20}{9} \frac{{\cal D} g_{\rm in}}{{\cal H}_0^2 \Omega_{\rm m0}} \Psi_0  \right) \nonumber \\
&\qquad + \frac 4 9 \frac{\dot {\cal D}^2}{({\cal H}_0^2 \Omega_{\rm m0})^2} \nab^{-2}  {\cal S}_{ij} + \frac 1 2 \pi_{2 ij_{\rm P}}
\Bigg] \,.
\end{align}
For the density contrast we obtain
\begin{align}
 \delta_{{\rm T}} &= \frac{2}{3 {\cal H}_0^2 \Omega_{\rm m0}}  {\cal D} \nab^2 \varphi_0
+ \frac{20 {\cal D} g_{\rm in}}{9{\cal H}_0^2 \Omega_{\rm m0}} \left[ \left( \frac 3 4 - a_{\rm nl} \right) (\nab \varphi_0)^2
    +  \left( 2 - a_{\rm nl} \right)  \varphi_0 \nab^2 \varphi_0 \right] \nonumber \\ 
  &\qquad+ \frac{2}{9 ({\cal H}_0^2 \Omega_{\rm m0})^2} \Bigg[ ({\cal D}^2+{\cal F}) (\nab^2\varphi_{0})^2 + 2{\cal D}^2 \varphi_{0,k} \nab^2 \varphi_{0}^{,k} + ({\cal D}^2 - {\cal F}) \varphi_{0,lm} \varphi_0^{,lm} \Bigg]   \,.
\end{align}

\subsection{The gauge generators}\label{summaryGauge}

\paragraph{From the Poisson gauge to the synchronous-comoving gauge} We use~(\ref{TdGi}) and then obtain the fluid trajectory as viewed from an observer in the Poisson gauge
\begin{align}
 x_{\rm P}^i(\tau,\fett{q}) &=  q^i  - \frac{2 {\cal D}}{3 \stuff} \varphi_{0}^{,i}
   + \frac{20 {\cal D} g_{\rm in}}{9 \stuff}  (a_{\rm nl} -2 ) \varphi_0\varphi_0^{,i}  - \frac{2}{3 \stuff} \left[  \frac{\dot {\cal D}^2}{\stuff}  + 5 {\cal D} g_{\rm in}  \right] \Theta_0^{,i} \nonumber \\
 &\qquad 
+ \frac{4}{9 {\cal H}_0^2\Omega_{\rm m0}} \left[ \frac{\dot {\cal D}^2}{{\cal H}_0^2\Omega_{\rm m0}} + 5 {\cal D} g_{\rm in} \right] {\cal R}^i  + \frac{4 {\cal F}}{9 (\stuff)^2} \Psi_0^{,i} \,,
\intertext{where $(\tau,\fett{q})$ are the Lagrangian coordinates. The relation between the temporal coordinates is}
  \eta_{\rm P}(\tau,\fett{q}) &= \tau - \frac{2 \dot {\cal D}}{3 \stuff} \varphi_0 
  + \frac{\dot {\cal D}}{3 \stuff} \left[  \frac{10}{3} g_{\rm in} \left( a_{\rm nl} - \frac 3 2 \right) \varphi_0^2 - 12 g \Theta_0 \right] \nonumber \\
&\qquad + \frac{2 \dot {\cal D} g}{3 \stuff} \varphi_0^2 + \frac{2 \dot {\cal D} {\cal D}}{9(\stuff)^2} (\nab \varphi_0)^2 + \frac{4 \dot {\cal F}}{9(\stuff)^2} \Psi_0 \,.
\end{align}

\paragraph{From the synchronous-comoving gauge to the total matter gauge}
We use~(\ref{TdG}) and then obtain the fluid trajectory as viewed from an observer in the total matter gauge
\begin{align}
 x_{\rm T}^i(\tau,\fett{q}) &=  q^i  - \frac{2 {\cal D}}{3 \stuff} \varphi_{0}^{,i}
   + \frac{20 {\cal D} g_{\rm in}}{9 \stuff}  (a_{\rm nl} -2 ) \varphi_0\varphi_0^{,i}  - \frac{10}{3 \stuff} {\cal D} g_{\rm in}  \Theta_0^{,i} \nonumber \\
 &\qquad 
+ \frac{20}{9 {\cal H}_0^2\Omega_{\rm m0}}  {\cal D} g_{\rm in} {\cal R}^i  + \frac{4 {\cal F}}{9 (\stuff)^2} \Psi_0^{,i} \,.
\end{align}
For the temporal coordinate we simply have  $\eta_{\rm T} = \tau$.

\paragraph{Comparison with the Newtonian trajectory}

In Newtonian gravity, the Newtonian trajectory is up to second order
\be \label{NLPTfinal}
  \bar x^i(\bar\tau,\bar{\fett{q}}) =  \bar q^i - \frac{2 {\cal D}}{3 \stuff} \varphi_{0}^{,i} + \frac{4 {\cal F}}{9 (\stuff)^2} \Psi_{ 0}^{,i} \,,
\ee
where we recall that bars denote Euclidean coordinates.

%%%%%%%%%%%%%%%%%%%%%%%%%%%%%%%%%%%%%%%%%%%%%%%%%%%%%%%%%%%%%%%%%%%%%%%%%%%%%%%%%%%%%%%%%%%%%%%%%%%%%%%%%%%%%%%%%%%%
\section{Conclusions}\label{conclude}
%%%%%%%%%%%%%%%%%%%%%%%%%%%%%%%%%%%%%%%%%%%%%%%%%%%%%%%%%%%%%%%%%%%%%%%%%%%%%%%%%%%%%%%%%%%%%%%%%%%%%%%%%%%%%%%%%%%%

We have considered relativistic perturbations for a collision-less and irrotational fluid up to second order in a $\Lambda$CDM Universe. In detail, we have determined the fastest growing mode solutions of all metric perturbations, the density and the velocity
for three common gauges, namely for the Poisson gauge, the total matter gauge 
and for the synchronous-comoving gauge. 
We considered growing mode initial conditions with primordial non-Gaussianity. 
First, we have confirmed the metric expressions of \cite{Bartolo:2005xa} 
in the Poisson gauge and corrected for some typos. 
We found major simplifications for the expressions in the Poisson gauge.
Let us emphasise that, because of these findings, all perturbations up to second order are fully described by only two time coefficients, which are identical with the first-order and the second-order growth of the 
Newtonian displacement field. Furthermore, our resulting solutions are very compact 
and come with simple physical interpretations.

Having determined the solutions in the Poisson gauge, we have performed a gauge transformation to the synchronous-comoving gauge. This gauge transformation can be understood, in terms of fluid mechanics, as a perturbative coordinate transformation 
from a specific Eulerian frame to the unique Lagrangian frame. Our results
in the synchronous-comoving gauge are, to our knowledge, 
entirely new, only the second-order density 
was known so far in the literature \cite{BMPR2010,Uggla:2013kya,Uggla:2014hva,BHMW14}.

Then, we have considered the gauge transformation from the synchronous-comoving gauge to the total matter gauge, which, as above, can be understood in terms of a coordinate transformation, here from the unique Lagrangian to another Eulerian frame.
This particular Eulerian frame is very important, as the resulting time coordinate in the total matter gauge is identical with the one in the synchronous-comoving gauge. That is, this particular Eulerian frame makes use of the proper time of the fluid particles. 
This choice of Eulerian frame is thus very convenient when relating GR results to Newtonian investigations, since the relativistic Eulerian-Lagrangian correspondence makes use of the same time coordinate. So to say, the problem of relating a GR description to a Newtonian world has been reduced from a 4D problem to a 3D problem.
Such considerations are for example useful when generating GR initial conditions for Newtonian $N$-body simulations (see \cite{FidlerEtAll}), the latter being one area of application of our paper.

Our results for the coordinate transformations in section~\ref{summaryGauge} dictate the relativistic trajectories of CDM particles in a $\Lambda$CDM Universe, and these results tell us explicitly how the Newtonian particle trajectories are deformed because of 
general relativistic effects. These GR corrections could be incorporated into $N$-body solvers with the aim to include GR effects in Newtonian $N$-body simulations effectively up to second order in relativistic perturbation theory.
Our results for the density and velocity in three common gauge choices 
have also direct outcomes for relating theory with observations of the large scale structure of the Universe. Specifically, measurements of the galaxy number density require both the knowledge of gauge effects (stemming from the choice of time-like hypersurfaces) and of GR effects due the fact that observations are performed on the past light-cone (e.g., observed photons are gravitationally distorted while passing through a clumpy Universe), the former being performed in the current paper and the latter being investigated elsewhere (see e.g., \cite{Bertacca:2014dra,Yoo:2014sfa}).

Finally, we have no reason to hide the pleasure that all our results are fairly simple. Of course, the reason for such simple expressions is hidden in the laborious work of solving the time differential equations in the Poisson gauge, which were not known so far.

\acknowledgments

We thank Marco Bruni, Kazuya Koyama, Roy Maartens, Sabino Matarrese, Claes Uggla, Obinna Umeh, John Wainwright, and David Wands for useful discussions and/or comments on the draft. We also thank the anonymous referee for her/his useful remarks.
EV thanks ``Fondazione Angelo della Riccia'' and the University of Portsmouth for financial support.  
CR acknowledges the support of the individual fellowship RA 2523/1-1 from the Deutsche Forschungsgemeinschaft.

%%%%%%%%%%%%%%%%%%%%%%%%%%%%%%%%%%%%%%%%%%%%%%%%%%%%%%%%%%%%%%%%%%%%%%%%%%%%%%%%%%%%%%%%%%%%%%%%%%%%%%%%%%%%%%%%%

\appendix

\section{Used notation}\label{app:notation}

In tab.\,\ref{tab:notation} we give an overview of the notation used in this work. For the case of an EdS universe, we have $\Omega_{\rm m} = 1$, $a=\eta^2$, ${\cal H}/2\eta$,
 $\stuff =4$, $g= g_{\rm in}=1$, ${\cal D}=\eta^2$, and ${\cal F}=(3/7) \eta^4$.

\renewcommand{\arraystretch}{1.2}

\begin{table}[t]

\hspace*{1.5cm}
\vskip1cm

 \begin{centering}

   \begin{tabular}{|lll|}
     \hline
     $g_{\mu\nu}$  &  metric ($\mu=0,1,2,3$)&  eq.\,(\ref{metricIIany}) \\
     $g_{ij}$ & spatial metric ($i=1,2,3$)  & eq.\,(\ref{metricICany}) \\
     $a(\eta)$  & cosmic scale factor  & eq.\,(\ref{scalef}) \\
     ${\cal H} = \dot a/a$  & conformal Hubble parameter & eq.\,(\ref{energy}) \\
     $\Omega_{\rm m}(\eta)$  & matter density & eq.\,(\ref{omegam}) \\  
     ${\cal D}(\eta)$  & structure growth of linear density fluctuations & eq.\,(\ref{growth}) \\
 ${\cal F}(\eta)$   & time coefficient of second-order Newtonian displacement & eq.\,(\ref{eqF})  \\
     $g(\eta)$  & growth suppression rate ($\equiv {\cal D}/a$) &  eq.\,(\ref{relphiphi_0}) \\
     $f= \dot {\cal D}/({\cal H}{\cal D}) $ & differential growth & eq.\,(\ref{psi2p}) \\
    % $g_{\rm in}$ & growth suppression rate  at initial time (deep in matter domination)  & eq.\,(\ref{})
     $\delta(\eta,\fett{x})$ & density contrast ($\delta = (\rho - \bar \rho)/\bar \rho$)   & eq.\,(\ref{def:delta}) \\
     $\varphi(\eta,\fett{x})= g(\eta) \varphi_0$ & linear gravitational potential
      at time $\eta$ & eq.\,(\ref{varphi}) \\
     $\varphi_0(\eta_0,\fett{x})$ & $\varphi$ at the present time $\eta_0$ & eq.\,(\ref{varphi}) \\
     $\varphi_{\rm in}= g_{\rm in} \varphi_0$ & linear gravitational potential at initial time & eq.\,(\ref{init}) \\ 
     $a_{\rm nl}= (3/5)f_{\rm nl}+1$  & primordial local non-Gaussianity & eq.\,(\ref{init}) \\
      $\nab^{-2}$  & inverse spatial Laplacian with Euclidean metric & eq.\,(\ref{chi2}) \\
   $D_{ij}$  &  longitudinal extraction operator ($D_{ij}=\partial_i \partial_j- (1/3) \nab^2 \delta_{ij}$)  & eq.\,(\ref{tens})  \\
     $\Psi_0(\eta_0,\fett{x})$ &  second-order ``Newtonian'' kernel for the displacement   &   eq.\,(\ref{Psi0})  \\
    $\Theta_0(\eta_0,\fett{x})$  & second-order ``GR'' kernel ($\nab^2 \Theta_0 = \Psi_0 - (1/3)(\nab\varphi_0)^2)$)  & eq.\,(\ref{Theta0}) \\
    ${\cal R}_i(\eta_0,\fett{x})$  &  second-order ``GR'' vector kernel   & eq.\,(\ref{transR}) \\
   ${\cal S}_{ij}(\eta_0,\fett{x})$  &  second-order ``GR'' tensor kernel  & eq.\,(\ref{tensorSij}) \\
   $\pi_{2ij_{\rm P}}(\eta,\fett{x})$ & secondary tensor perturbations in the Poisson gauge  & eq.\,(\ref{GW}) \\
   $\psi$ & scalar perturbation in $g_{00}$  & eq.\,(\ref{metricIIany}) \\
   $B$  & scalar perturbation  in $g_{0i}$ & eq.\,(\ref{vec}) \\
    $\omega_i$ & vector perturbation in $g_{0i}$ & eq.\,(\ref{vec}) \\
   $\phi$  & scalar perturbation in the trace of $g_{ij}$  & eq.\,(\ref{tens}) \\
   $E$  & scalar perturbation in the trace-less part of $g_{ij}$  & eq.\,(\ref{tens}) \\
   $F_i$  & vector perturbation in $g_{ij}$  & eq.\,(\ref{tens}) \\
   $\chi_{ij}$  & tensor perturbation in $g_{ij}$  & eq.\,(\ref{tens}) \\
   $u^\mu= (\delta^\mu_0+v^\mu)/a$  & 4-velocity ($u^\mu u^\nu g_{\mu\nu} =-1$) with perturbation $v^\mu$ & eq.\,(\ref{umu}) \\
   $v^0$ & perturbation in the time-component of $u^\mu$ & eq.\,(\ref{v0any}) \\
   $v_{i} = v_{,i} + w_{i}$ & (longitudinal and transverse) spatial velocity perturbation & eq.\,(\ref{velocity}) \\
   $\xi^\mu = (\xi^0, \xi^i)$ & gauge generator  & eq.\,(\ref{TdG}) \\
   $\alpha = \xi^0$ & time component of the gauge generator & eq.\,(\ref{xidecomp}) \\
   $\xi^i = \beta^{,i} + d^i$ & (longitudinal and transverse part of) spatial gauge 
     generator & eq.\,(\ref{xidecomp}) \\
     \hline
    \end{tabular}

 \end{centering}

\caption{Used notation in this work. %We denote the (comoving) Lagrangian coordinates with $(\tau,\fett{q})$, and the (comoving) Eulerian coordinates with $(\eta,\fett{x})$.
}
\label{tab:notation}

%\vskip1cm

\end{table}

%\newpage

%%%%%%%%%%%%%%%%%%%%%%%%%%%%%%%%%
\section{Lie derivatives} \label{lieapp}
The Lie derivatives $\mathcal{L}_{\xi}$ and $\mathcal{L}^2_{\xi}$ along the vector field $\xi^\mu$ have the following expressions: 
\begin{itemize}
\item scalars
\begin{eqnarray}
\mathcal{L}_{\xi} f &=&f_{,\nu}\xi^\nu \\
\mathcal{L}^2_{\xi} f &=&f_{,\omega\ss}\xi^\omega\xi^\ss+f_{,\omega}\xi^\omega_{,\nu}\xi^\nu+f_{,\ss}\zeta^\ss
\end{eqnarray}
\item vectors
\begin{eqnarray}
\mathcal{L}_{\xi} v^\a &=&v^\a_{,\nu}\xi^\nu-\xi^\a_{,\nu}v^\nu  \\ 
\mathcal{L}^2_{\xi} v^\a &=& v^\a_{,\nu\rho}\xi^\nu\xi^\rho+v^\a_{,\nu}\xi^\nu_{,\rho}\xi^\rho-2\xi^\a_{,\rho}v^\rho_{,\nu}\xi^\nu-\xi^\a_{,\nu\rho}\xi^\nu v^\rho +\xi^\a_{,\rho}\xi^\rho_{,\nu}v^\nu     
\end{eqnarray}
\item 1-forms
\begin{eqnarray}
\mathcal{L}_{\xi} \omega_\a &=&\omega_{\a,\nu}\xi^\nu+\xi^\nu_{,\a}\omega_\nu  \\
\mathcal{L}^2_{\xi} \omega_\a &=&\omega_{\a,\nu\rho}\xi^\nu\xi^\rho+\omega_{\a,\nu}\xi^\nu_{,\rho}\xi^\rho+2\xi^\rho_{,\a}\omega_{\rho,\nu}\xi^\nu+\xi^\rho_{,\nu\a}\xi^\nu \omega_\rho +\xi^\nu_{,\rho}\xi^\rho_{,\a}\omega_\nu \,.    
\end{eqnarray}
\item metric tensor
\begin{eqnarray}
\mathcal{L}_{\xi} g_{\mu\nu} &=& g_{\mu\nu,\sigma}\xi^\sigma+\xi^\sigma_{,\nu}g_{\mu\sigma}+\xi^\sigma_{,\mu}g_{\nu\sigma}\\
\mathcal{L}^2_{\xi} g_{\mu\nu} &=&2\xi^\omega_{,\mu}g_{\omega\nu,\sigma}\xi^\sigma+\xi^\omega_{,\mu}\xi^\sigma_{,\omega}g_{\sigma\nu}+2\xi^\omega_{\mu}\xi^\sigma_{,\nu}g_{\omega\sigma}+2\xi^\omega_{,\nu}g_{\omega\mu,\sigma}\xi^\sigma \nonumber\\
&& +\xi^\omega_{,\nu}\xi^\sigma_{,\omega}g_{\mu\sigma}+g_{\mu\nu,\sigma\omega}\xi^\sigma\xi^\omega+\xi^\sigma_{,\mu\omega}g_{\sigma\nu}\xi^\omega+\xi^\sigma_{,\nu\omega}g_{\mu\sigma}\xi^\omega \,. \label{lie2g}
\end{eqnarray}
\end{itemize}

%%%%%%%%%%%%%%%%%%%%%%%%%%%
\section{First-order squared terms arising at second order}\label{app:FOsquaredterms}
The contributions from products of first-order terms to the second-order gauge transformations in section~\ref{secGauge} are given by\footnote{Note that in the last line of eq.~\ref{defO} we have corrected a typo of eq.\,(6.27) of \cite{malik&wands}. }
\begin{align}
\Pi &\equiv \alpha_1 \left[ \ddot \alpha_1 + 5 {\cal H} \dot \alpha_1 + (\dot {\cal H} + 2 {\cal H}^2) \alpha_1 + 4 {\cal H} \psi_1 + 2 \dot \psi_1 \right] \nonumber \\
&\qquad+ 2 \dot \alpha_1 (\dot \alpha_1 +  2 \psi_1) + \xi^k_{1} (\dot \alpha_1 + {\cal H} \alpha_1 + 2 \psi_1)_{,k} + \dot \xi_{1}^k \left[\alpha_{1, k} - 2 B_{1k} - \dot \xi_{1k}\right] \label{defP}, \\ %ok with MMB and MW (where it is not defined)
\Sigma_{i} &\equiv 2 \left[ (2{\cal H} B_{1i} + \dot B_{1i}) \alpha_{1}+ B_{1i, k} \xi^k_1 - 2 \psi_1 \alpha_{1,i} + B_{1k} \xi^k_{1,i} + B_{1i} \dot \alpha_1 + 2 C_{1ik} \dot \xi_1^k \right] \nonumber \\
&\qquad + 4 {\cal H} \alpha_1 (\dot \xi_{1i} - \alpha_{1,i}) + \dot \alpha_1 (\dot \xi_{1i} - 3 \alpha_{1,i}) + \alpha_1 (\ddot \xi_{1i} - \dot \alpha_{1,i}) \nonumber\\
 &\qquad + \dot \xi_1^k (\xi_{1i,k} + 2 \xi_{1k,i}) + \xi^k_1 (\dot \xi_{1i,k} - \alpha_{1,ik}) - \alpha_{1,k} \xi^k_{1,i}\,, \label{defS}\\ % =6.49MW see 6.51MW for \Sigma^k_{,k}\\ 
\Omega^i &\equiv \dot{\xi}^i_{1}\left(2\psi_1+\dot{\alpha}_1+2{\cal H}\alpha_1\right)-\alpha_1\ddot{\xi}^i_{1}-\xi_1^k\dot{\xi}^i_{1,k}+\dot{\xi}_1^{k}\xi^i_{1,k}
\nonumber\\
&\qquad -2\alpha_1\left({\cal H} v^i_{1}  - \dot{v}^i_{1} \right)+2v^i_{1,k}\xi_1^k-2v_1^k\xi^i_{1,k}
\,,\label{defO}\\% =typo in 6.27MW (term with \dot{v} has the wrong sign) =first order product 3.32MMB
\intertext{and}
\Upsilon_{ij} &\equiv 2 \left[ (\dot {\cal H}+2{\cal H}^2) \alpha_1^2 + {\cal H}(\alpha_1 \dot \alpha_1 + \alpha_{1,k} \xi_1^k)\right] \delta_{ij} \nonumber \\
&\qquad - 2 \alpha_{1,i} \alpha_{1,j} + 4 \alpha_1 (\dot C_{1 ij} + 2 {\cal H} C_{1 ij} ) + 4 B_{1 (i} \alpha_{,j)} \nonumber \\
&\qquad +4 \left[C_{1ij,k} \xi_1^k + 2 C_{1 k ( i} \xi^k_{1, j)} \right] + 8 {\cal H} \alpha_1 \xi_{1 (i, j)} + 2 \xi_{1 k, i} \xi_{1,j}^k \nonumber \\
&\qquad + 2\alpha_1 \dot \xi_{1 (i,j)} + 2 \xi_{1 (i, j)k} \xi_1^k  + 2\xi_{1 (i, k} \xi^k_{1, j)}+ 2 \dot \xi_{1 (i} \alpha_{1, j)}  \, . \label{defU} %= 6.54MW and ok con MMB, see 6.56MW for \Upsilon^k_k. note that \Upsilon^k_k=-1/6 * (product of I order terms in 3.26MMB)
\end{align}
In the next two subsections we give the explicit expressions for our specific transformations.

\paragraph{From the Poisson to the synchronous-comoving gauge}\label{app:FOsquaredtermsPS}
The second-order contributions which arrive solely from first-order squared terms are in the case Poisson gauge to synchronous-comoving gauge
\begin{align}
  \Pi_{{}_\PS} &=  \alpha_{1_\PS} \dot g \varphi_0 - 2 g^2 \varphi_0^2 + \beta_{1_\PS}^{\ , k} g \varphi_{0,k} = \frac{\dot {\cal D}}{a} \alpha_{1_\PS} \varphi_0 + \frac 5 3 \varphi_{\rm in} g \varphi_0 - 3 g^2 \varphi_0^2  + \beta_{1_\PS}^{\ , k} g \varphi_{0,k}  \nonumber \\
  &= \left( \frac 5 3 g g_{\rm in} -\frac{2 \dot {\cal D}^2}{3a \stuff} -3g^2 \right)  \varphi_0^2 - \frac{2{\cal D}^2}{3a \stuff} \varphi_{0,l} \varphi_0^{,l} \,, \\
\Sigma_{i_\PS} &= -6 g \varphi_{0} \alpha_{1_\PS,i} + 2 {\cal H} \alpha_{1_\PS} \alpha_{1_\PS,i}  +2 \alpha_{1_\PS,k} \beta_{1_\PS,i}^{,k}  \nonumber \\ 
 &= -4 g \varphi_0 \alpha_{1_\PS,i} - \frac{10}{3} \varphi_{\rm in} \alpha_{1_\PS,i} + 2\alpha_{1_\PS,k} \beta^{,k}_{1_\PS,i} \nonumber  \\
  &= \frac{8}{3} \frac{{\cal D} \dot {\cal D}}{a {\cal H}_0^2 \Omega_{\rm m0}} \varphi_{0} \varphi_{0,i}
  +\frac{20}{9} \frac{g_{\rm in} \dot {\cal D}}{{\cal H}_0^2 \Omega_{\rm m0}} \varphi_0 \varphi_{0,i}  +\frac{8}{9} \frac{{\cal D} \dot {\cal D}}{({\cal H}_0^2 \Omega_{\rm m0})^2} \varphi_{0,k}\varphi_{0,i}^{,k}  \,, \label{sigmaS}\\
\Omega^i_{\PS} &= 2g \varphi_0 \alpha^{,i}_{1_\PS} +2 \alpha_{1_\PS} \dot \alpha^{,i}_{1_\PS}  = \frac{2 \dot {\cal D}}{3 \stuff} \left[ \frac{2 \ddot {\cal D}}{3\stuff} - g  \right] \partial^i \varphi_0^2
 \nonumber \\ &= - \frac{4 {\cal H} \dot {\cal D}^2}{9\stuff^2} \partial^i \varphi_0^2   
  =   \frac{2 \dot {\cal D}}{3 \stuff} \left[  g - \frac 5 3 g_{\rm in} \right] \partial^i \varphi_0^2 \,, \label{omegaS}
\\
\Upsilon_{ij_\PS} &=  \left[ 2 \frac{\ddot a}{a} \alpha_{1_\PS}^2 -6{\cal H} \alpha_{1_\PS} g \varphi_{0}  +2 {\cal H} \alpha_{1_\PS,k} \beta_{1_\PS}^{,k} - 4\alpha_{1_\PS} \frac{\dot {\cal D}}{a}  \varphi_0 -4 g \varphi_{0,k} \beta_{1_\PS}^{,k} \right] \delta_{ij} \nonumber  \\ 
  &\qquad  - 8g \varphi_0 \beta_{1_\PS,ij} + 8 {\cal H} \alpha_{1_\PS} \beta_{1_\PS,ij} + 4 \beta_{1_\PS,ki}\beta_{1_\PS,j}^{,k} + 2 \alpha_{1_\PS} \alpha_{1_\PS,ij} + 2 \beta_{1_\PS,ijk} \beta^{,k}_{1_\PS} \nonumber \\
&= \left[ \left( \frac{100}{9} g_{\rm in}^2 - \frac{10}{3} g g_{\rm in} - 2g^2 + \frac 4 3 \frac{\dot {\cal D}^2}{a{\cal H}_0^2 \Omega_{\rm m0}} \right) \varphi_0^2 
    +\frac{20}{9} \frac{g_{\rm in} {\cal D}}{{\cal H}_0^2 \Omega_{\rm m0}} \varphi_{0,k} \varphi_{0}^{,k}
  + \frac 4 3 \frac{{\cal D}^2}{a {\cal H}_0^2 \Omega_{\rm m0}} \varphi_{0,k} \varphi_{0}^{,k} \right] \delta_{ij} \nonumber \\
  &+ \frac{80}{9} \frac{g_{\rm in} {\cal D}}{{\cal H}_0^2 \Omega_{\rm m0}} \varphi_{0} \varphi_{0,ij} 
  + \frac 8 9 \frac{\dot {\cal D}^2}{({\cal H}_0^2 \Omega_{\rm m0})^2} \varphi_0 \varphi_{0,ij}
  + \frac{16}{9} \frac{{\cal D}^2}{({\cal H}_0^2 \Omega_{\rm m0})^2} \varphi_{0,ik} \varphi_{0,j}^{,k}
 + \frac 8 9 \frac{{\cal D}^2}{({\cal H}_0^2 \Omega_{\rm m0})^2} \varphi_{0,kij} \varphi_{0}^{,k}  \label{upsilonS} 
\end{align}
%%%%%%%%%%%%%
Here we have made use of eq.\,(\ref{eq:constIN}).
It is worthwhile to derive the trace of $\Upsilon_{ij_\PS}$ explicitly:
\begin{align}
  \Upsilon^k_{k_\PS} & = \left( \frac{100}{3} g_{\rm in}^2 - 10 g g_{\rm in} - 6g^2 + 4  \frac{\dot {\cal D}^2}{a{\cal H}_0^2 \Omega_{\rm m0}} \right) \varphi_0^2  \nonumber \\
  &+ \frac{20}{3} \frac{g_{\rm in} {\cal D}}{{\cal H}_0^2 \Omega_{\rm m0}} \varphi_{0,k} \varphi_{0}^{,k}
  + 4 \frac{{\cal D}^2}{a {\cal H}_0^2 \Omega_{\rm m0}} \varphi_{0,k} \varphi_{0}^{,k}
  + \frac{80}{9} \frac{g_{\rm in} {\cal D}}{{\cal H}_0^2 \Omega_{\rm m0}} \varphi_{0} \nab^2 \varphi_0  \nonumber \\
  &+ \frac 8 9 \frac{\dot {\cal D}^2}{({\cal H}_0^2 \Omega_{\rm m0})^2} \varphi_0 \nab^2 \varphi_{0}
  + \frac{16}{9} \frac{{\cal D}^2}{({\cal H}_0^2 \Omega_{\rm m0})^2} \varphi_{0,kl} \varphi_{0}^{,kl}
  + \frac 8 9 \frac{{\cal D}^2}{({\cal H}_0^2 \Omega_{\rm m0})^2} \varphi_{0,k} \nab^2\varphi_{0}^{,k} \,.
\end{align}

\paragraph{From the synchronous-comoving to the total matter gauge} \label{app:FOsquaredtermsST}
Since there is no time transformation involved, the second-order contributions from first-order squared terms reduce to
\begin{align}
\Pi_{{}_\ST} &= - \dot \xi_{1_\ST}^k  \dot \xi_{1 k_\ST}
  =  - \frac 4 9 \frac{\dot {\cal D}^2}{({\cal H}_0^2 \Omega_{\rm m0})^2} \varphi_{0,k} \varphi_{0}^{,k} \label{Px}, \\ 
\Sigma_{i_{\ST}} &= 4 C_{1ik_{\rm S}} \dot \xi_{1_\ST}^k+ \dot \xi_{1_\ST}^k (\xi_{1 i_\ST,k} + 2 \xi_{1k_\ST,i}) + \xi^k_{1_\ST} \dot \xi_{1 i_\ST,k}=  - \frac{40}{9} \frac{\dot {\cal D} g_{\rm in}}{{\cal H}_0^2 \Omega_{\rm m0}}  \varphi_{0} \varphi_{0,i} \,, \label{Sx}\\  
\Omega^i_{{\ST}} &= 0 \,,\label{Ox}\\
\intertext{ 
and}
\Upsilon_{ij_{\ST}} &= 4 \left[C_{1ij_{\rm S},k} \xi_{1_\ST}^k + 2 C_{1 k ( i_{\rm S}} \xi^k_{1_\ST, j)} \right] +  2 \xi_{1 k_\ST, i} \xi_{1_\ST,j}^k + 2 \xi_{1_\ST (i, j)k} \xi_1^k  + 2\xi_{1_\ST (i, k} \xi^k_{1_\ST, j)}  \nonumber \\
 &= -\frac{40}{9} \frac{{\cal D} g_{\rm in}}{{\cal H}_0^2 \Omega_{\rm m0}} \varphi_{0,k} \varphi_{0}^{,k} \delta_{ij} - \frac{80}{9} \frac{{\cal D} g_{\rm in}}{{\cal H}_0^2 \Omega_{\rm m0}} \varphi_{0} \varphi_{0,ij}
  - \frac 8 9 \frac{{\cal D}^2}{({\cal H}_0^2 \Omega_{\rm m0})^2} \left( \frac 1 2 (\varphi_{0,k}\varphi_{0}^{,k})_{,ij} + \varphi_{0,ik} \varphi_{0,j}^{,k}  \right)
  \label{Ux}
\end{align}
where for the second equality we have substituted the various expressions from the metric in the synchronous-comoving gauge and from the first-order transformations.
It is worthwile to derive the trace of $\Upsilon_{ij_\ST}$ explicitly: 
\begin{align}
  \Upsilon^k_{k_\ST} &= 
 -\frac{40}{3} \frac{{\cal D} g_{\rm in}}{{\cal H}_0^2 \Omega_{\rm m0}} \varphi_{0,k} \varphi_{0}^{,k}  - \frac{80}{9} \frac{{\cal D} g_{\rm in}}{{\cal H}_0^2 \Omega_{\rm m0}} \varphi_{0} \nab^2 \varphi_{0}
  - \frac 8 9 \frac{{\cal D}^2}{({\cal H}_0^2 \Omega_{\rm m0})^2} \left( \frac 1 2 \nab^2 (\varphi_{0,k}\varphi_{0}^{,k}) + \varphi_{0,kl} \varphi_{0}^{,kl}  \right) \,.
\end{align}

%%%%%%%%%%%%%%%%%%%%%%%%%%%%%%

%%%%%%%%%%%%%%%%%%%%%%%%%%%%%%

\end{document}